\documentclass[manuscript]{aastex}
\usepackage{subfig}
\pdfoutput=1

\newcommand{\myemail}{lisa.harvey-smith@csiro.au}
\def\zetaoph{$\zeta$~Oph}

\def\alphacam{$\alpha$~Cam}
\def\lambdaori{$\lambda$~Ori}
\def\chiper{$\chi$~Per}

\def\halpha{H$\alpha$}

\newcommand\HII{H\,{\sc ii}}

\shorttitle{Magnetic fields in Galactic \HII\ regions}
\shortauthors{Harvey-Smith et al.}

\begin{document}

\title{Magnetic Fields in Large Diameter \HII\ Regions Revealed by the Faraday Rotation of Compact Extragalactic Radio Sources}

\author{L. Harvey-Smith,\altaffilmark{1,2} G. J. Madsen \altaffilmark{1} and B. M. Gaensler\altaffilmark{1,3}}

\altaffiltext{1}{Sydney Institute for Astronomy (SIfA), School of Physics,
The University of Sydney, NSW 2006, Australia}
\altaffiltext{2}{Present Address: CSIRO Astronomy and Space Science, Australia Telescope National Facility, PO Box 76, Epping, NSW 2121, Australia; \myemail }
\altaffiltext{3}{Australian Research Council Federation Fellow}

\begin{abstract}
We present a study of the line-of-sight magnetic fields in five large-diameter Galactic \HII\ regions. Using the Faraday rotation of background polarized radio sources, as well as dust-corrected \halpha\ surface brightness as a probe of electron density, we estimated the strength and orientation of the magnetic field along 93 individual sight-lines through the \HII\ regions. Each of the \HII\ regions displayed a coherent magnetic field.  The magnetic field strength (line-of-sight component) in the regions ranges from 2 to 6 $\mu$G, which is similar to the typical magnetic field strength in the diffuse interstellar medium.  We investigated the relationship between magnetic field strength and electron density in the 5 \HII\ regions. The slope of magnetic field vs. density in the low-density regime ($0.8 < $ $n_e$ $ < 30$ cm$^{-3}$) is very slightly above zero. We also calculated the ratio of thermal to magnetic pressure, $\beta_{th}$, for each data point, which fell in the range 1.01 $< \beta_{th} <$ 25. Finally, we studied the orientation of the magnetic field in the solar neighborhood (d $<$ 1.1~kpc) using our data from 5 \HII\ regions along with existing measurements of the  line-of-sight magnetic field strength from polarized pulsars whose distances have been determined from their annual parallax. We identify a net direction for the magnetic field in the solar neighborhood, but find no evidence for a preferred vertical direction of the magnetic field above or below the Galactic plane.
\end{abstract}


\section{Introduction}
\label{intro}

Understanding the interactions between neutral and ionized matter and magnetic fields is crucial in testing models of star-formation in the Galaxy. The expectation from ideal magnetohydrodynamic flux freezing is that the total magnetic field strength, $B$, is proportional to $n^{\kappa}$, where $n$ is the gas number density and 0 $<$ $\kappa$ $<$ 1, depending on the geometry of bulk gas motions relative to the magnetic field lines \citep{crutcher99}. To date, observational estimates of magnetic field strengths in a large number of Galactic molecular clouds suggest that $\kappa \sim 0.4 - 0.6$ in regions with volume densities between 10$^3-$10$^5$~cm$^{-3}$ \citep{th86,vallee95,crutcher07,cwh+10}. This range of $\kappa$ probably represents an upper limit, as processes such as fast ambipolar diffusion (ion-neutral drift) may also play an important role within dense, turbulent regions \citep{zweibel02}. In contrast, regions with lower gas densities (in the range of 10$^{-1}$$-$10$^3$~cm$^{-3}$) appear to have magnetic fields of a fairly constant $\sim$5 $-$ 10~$\mu$G \citep{crutcher07}. A simple interpretation is that, in the absence of strong turbulence, low-density gas in the ISM is free to move along magnetic field lines without inciting a change in magnetic field strength \citep{fm93, pn99, kimo06}. The constancy in the strength of the magnetic field at low densities has been predicted by models \citep{th86,fm93, hsh09}, but the mechanism by which material accumulates without increasing the magnetic flux is not clearly understood. Numerical simulations by \citet{heitsch+07,hsh09} have shown that gas flowing parallel to magnetic field lines may generate strong turbulent instabilities which, through the depletion of magnetic flux by reconnection or magnetic flux dissipation, may actually \emph{enhance} the correlation between $B$ and $n$ rather than destroying it. A major obstacle to understanding the process of large-scale star formation is the lack of reliable measurements of magnetic fields at relatively low gas densities \citep{th86,crutcher07}. Particularly valuable are \emph{in situ} measurements of $B$ and $n$ across a range of gas densities. 

Measurements of magnetic fields at a range of low gas densities also enable us to evaluate the magnetic and thermal pressures, which are useful inputs to numerical simulations of galaxy structure. These models often assume that the weight of the gas layer in the Galactic disk is counteracted by the total pressure of the disk gas \citep{po07}. \citet{oml10} presented a theory of star formation in a multi-phase interstellar medium (ISM) that is self-regulated by stellar heating.  The ratio of total to thermal pressure in low-density gas is a vital input to this model, but is poorly constrained by observations, and estimates quoted by \citet{oml10} vary by an order of magnitude.  We can assist these models by providing measurements of the thermal and magnetic pressures in the ISM over a range of densities.

Calculating the magnetic pressure requires a robust in-situ measurement of magnetic field strength, which is a difficult task. An excellent probe of magnetic fields is Faraday rotation of extragalactic radio sources, provided that the Faraday rotation due to the magnetized plasma in the region under scrutiny can be separated from the Faraday rotation due to the remainder of the Galactic interstellar medium along the line of sight. By focusing on discrete regions of relatively high electron density, i.e. \HII\ regions, it is possible to detect the Faraday rotation `signal' due to these localized regions above the `noise' provided by Galactic foreground and background. With this in mind, we here present a study of a small sample of large-diameter \HII\ regions, which we use to derive the magnetic field strength and $\beta_{th}$ across a range of gas densities.


In this experiment, we have used Faraday Rotation of background polarized radio sources and previously published maps of \halpha\ surface brightness to measure the magnetic field and gas density in 5 \HII\ regions that lie within 1.1~kpc of the Sun. By focusing on discrete regions of relatively high electron density and calibrating against the mean Faraday rotation along similar sightlines, it becomes possible to isolate the Faraday rotation due to the \HII\ region. The technique we use is based upon the foundations laid by \citet{hc80} and \citet{hct81}, who estimated magnetic field strengths in \HII\ regions using the Faraday rotation of just one or two background polarized sources. With a much larger number of background polarized sources (93 in total) to choose from, we are able to significantly advance this technique to accurately measure the line-of-sight component of the magnetic field strength in 5 discrete ionized volumes of the Galaxy.
 
This paper is organised as follows. In Section 2 we introduce our method for measuring $B_{||}$, $n$ and $\beta_{th}$ in \HII\ regions. Section 3 contains a description of our source selection process and our model for the \HII\ regions. Section 4 describes our results, including maps of the \HII\ regions, plots of rotation measure vs. \halpha\ surface brightness and plots of $B$ vs. $n$. In Section 5, we compare our work with previous results, examine the relationship between magnetic field strength and gas density in our \HII\ regions, and estimate the corresponding ratio of magnetic to thermal pressure. We comment on the orientation of the large-scale magnetic field in the local Galaxy, both above and below the plane. Our conclusions are presented in Section 6.

\section{Experimental Details}
\label{theory}

\subsection{Probes of magnetic field and electron density in {\sc Hii} regions}
\label{probes}
\HII\ regions are Faraday screens, i.e. the magnetized ionized plasma causes the electric field vector of a linearly polarized wave traversing the region to rotate by an angle that is proportional to the square of the observed wavelength, $\lambda^2$. Rotation Measure, RM, is the scaling factor between $\lambda^2$ and the change in polarization position angle, $\chi$, due to Faraday rotation, i.e. $\Delta\chi$ = RM $\lambda^2$. Quantitatively, the RM of an \HII\ region is described by Equation~(\ref{equation0}):

\begin{equation}
\label{equation0}
{\rm RM} = 0.81 \int n_e  B_{||} ~ \textit{dl}  ~{\rm rad~m^{-2}} 
\end{equation}

\noindent where $n_e$ is the density in cm$^{-3}$  of free electrons in the Faraday rotating medium, $B_{||}$ is the parallel component of the magnetic field in $\mu$G and the integral along a line element $dl$ runs from the source to the observer in parsecs. Positive and negative RMs indicate average magnetic fields pointed towards and away from the observer, respectively. 

The rotation measure of polarized radiation from a background source can be used as a probe of the electron-weighted parallel component of the magnetic field strength along a single sightline through an \HII\ region. The most straightforward way to infer $B_{||}$ from RM is to use a background pulsar. The dispersion measure, DM, of a pulsed electromagnetic signal from a pulsar is given by ${\rm DM} = \int_{obs}^{psr} {n_e}  ~ \textit{dl}  ~{\rm cm^{-3}~pc}$, where the integral limits are the positions of the observer and the pulsar. The average magnetic field strength along the line of sight to the pulsar is simply given by $B_{||} = \frac{\rm{RM}}{0.81\rm{DM}}$. However, the measured DM represents the sum of free electrons inside the \HII\ region and in other gas along the same sightline. Therefore the magnetic field \emph{within the \HII\ region} may only be inferred if a foreground subtraction can be made for DM using other pulsars at a similar distances in the same region of sky that do not intersect the \HII\ region. Similarly, a correction for Faraday rotation in the Galaxy is made using RMs of polarized extragalactic sources that lie close to, but outside the \HII\ region boundary. Unfortunately, the number of pulsars for which distances have been accurately determined is sufficiently low that this situation does not commonly arise.

There is another source of information on rotation measure aside from pulsars $-$ namely the increasingly large sample of unresolved polarized extragalactic sources. All extragalactic sources lie at essentially the same distance (i.e. at infinity) relative to the magneto-ionized material in the Galaxy, so reliable corrections for the Galactic RM can be made using an ensemble of extragalactic sources in a surrounding region of the sky. We use the RMs of polarized background extragalactic point sources to probe the electron-weighted magnetic field strength of several different paths through \HII\ regions. Our source of data on Faraday rotation is a recently published catalog of the RMs of 37,543 polarized radio sources  \citep{tss09} derived from the NRAO VLA Sky Survey \citep[NVSS;][]{ccg+98} at 1.4~GHz. 

\subsection{Source Selection}
\label{hii}
To identify a suitable source sample for this study, we searched maps of the \halpha\ sky for prominent \HII\ regions. To do this, we used maps of \halpha\ surface brightness I$_{H\alpha}$, with 6$^{\prime}$ resolution \citep{fink03} compiled from the WHAM \citep{hrt+03}, SHASSA \citep{gmr+01} and VTSS \citep{dst98} surveys. Our search was restricted to regions at least 5$^{\circ}$ away from the Galactic plane, as the RM signal close to the plane is dominated by the dense magnetized plasma in the Galaxy and dust extinction maps used to correct emission measures are unreliable \citep{sfd98}. We identified 5 \HII\ regions which lie at Galactic latitudes $|b|$ $>  $5$^{\circ}$, which have a peak I$_{H\alpha}$ at least 3 times the root mean square surface brightness of the surrounding region and have at least 4 background RM measurements in the catalogue of \citet{tss09}. These were Sh 2-27, Sh 2-264, Sivan 3, Sh 2-171 (NGC 7822) and Sh 2-220 (The California Nebula). Table~\ref{table1} shows the basic parameters for each region adopted in our analysis, including the number of RMs from \citet{tss09} lying behind each region, the diameter of the \HII\ region, and properties of the ionizing star: distance, reddening and spectral type. For each \HII\ region, we defined the boundary as the line within the regions of sky shown in Figure \ref{figure1} for which I$_{H\alpha} = 3~\sigma_{RMS}$, where $\sigma_{RMS}$ is the root mean square of I$_{H\alpha}$ outside the boundary (see details in Table~\ref{table2}). 

\subsection{Calculating $B_{||}$ for our \HII\ region sample}
\label{calcb}

In Section \ref{probes} we stated that RM probes the electron-weighted magnetic field strength along the line-of-sight. Therefore, by estimating the electron density we can disentangle $B_{||}$ from RM in \HII\ regions. We do this by measuring the Emission Measure, EM which is defined as ${\rm EM} =  \int_0^{\infty} n_e^2~ \textit{dl}  ~{\rm cm^{-6}}$. There are several observational probes of EM, the most common being the $n$ = 3 $-$ 2 Balmer (\halpha) recombination line of atomic hydrogen at 6563\AA. In order to determine the EM corresponding to each RM from the \citet{tss09} catalogue, we used the \citet{fink03} compilations of I$_{H\alpha}$ to estimate the electron density corresponding to each measured value of RM.

The EM along a path is related to I$_{H\alpha}$ \citep{rey88} in the following way:

\begin{equation}
\label{eq1}
{\rm EM} = 2.75~ \left(\frac{T_e}{10^{4}{\rm K}}\right)^{0.9}  \left(\frac{I_{H\alpha}}{{\rm R}}\right) e^{\tau}
 ~{\rm cm^{-6}~pc}.
\end{equation}

\noindent Here, $T_e$ is the electron temperature and $I_{H\alpha}$ is in rayleighs, where 1~R$=$10$^6$/4$\pi$ photons~s$^{-1}$~cm$^{-2}$~sr$^{-1}$. A correction term for dust extinction, $e^{\tau}$,  where the optical depth $\tau = 2.44 \times E_{B-V}$ \citep{fink03}, is included to extract the intrinsic $I_{H\alpha}$ from the measured value. This assumes that all the dust is in front of the source, rather than mixed in the source or behind the source. Several features in the dust maps towards our five \HII\ regions were also visible in the \halpha\ maps, indicating that much of the dust is in front of the \HII\ regions. We therefore applied the maximum possible correction to I$_{H\alpha}$, assuming that all the dust is situated between the \HII\ region and the Earth. Our dust corrected EMs should therefore be regarded as upper limits.

We adopt the following formalism to allow for possible variations in electron density along any line of sight through an \HII\ region. Our model assumes that along a single line of sight with depth $L$ ~pc, the ionized material is composed of discrete clumps. These clumps occupy a total length $fL$, where $f$ is the volume filling factor. Where ionized material is present (i.e. within a clump), the electron density is $n_e = n_0$. Outside the clumps we assume that $n_e = 0$.

The electron density in a clump is given by

\begin{equation}
\label{eq2} 
n_0 = \sqrt \frac{\rm{EM}}{f L}  ~{\rm cm^{-3}}, 
\end{equation}

and the average density along a sightline is $\langle n \rangle$ = $f n_0$.

\noindent In this case the strength of the parallel component of the magnetic field in the \HII\ region is 

\begin{equation}
\label{eq3} 
 B_{||} = \frac{{\rm RM}}{0.81 \sqrt{\rm{EM}} \sqrt{f L}}  ~\mu G,
\end{equation}

\noindent assuming that $n_0$ and $B_{||}$ are uncorrelated along a single sight-line. If fluctuations in the electron density and magnetic field are 
positively (negatively) correlated, then the use of Equation~(\ref{eq3}) will lead to an 
overestimate (underestimate) of the magnetic field strength \citep[see][]{bss+03}. We will discuss this point further in Section~5.3.

To evaluate Equation~(\ref{eq3}), RM and EM are determined from observations. The total path through the \HII\ region,  $L$, can be estimated from the geometry of the region and $f$ may be assumed to within some reasonable physical limits. Therefore, using Equations~(\ref{eq2}) and (\ref{eq3}) we can estimate  $n_0$ and $B_{||}$ at the position of each RM measurement. 

Once these quantities are known we can calculate the magnetic pressure $P_{mag}$, thermal pressure $P_{ther}$ and hence the plasma beta ($\beta_{th}$= $P_{ther}/P_{mag}$) for each data point as follows. Using the statistical approximation $B_{tot} \sim \sqrt 3~B_{||}$ and the observational estimate $\langle B_{uniform}/B_{tot} \rangle \le \sqrt 2$ \citep{beck01} we first estimate the total magnetic field strength $B_{tot} \sim \sqrt 6 B_{||}$ from the measured magnetic field strength, $B_{||}$ for each data point. The magnetic pressure $P_{mag}$= $B_{tot}^2$/8$\pi$ and the thermal pressure, $P_{ther}$ = $2n_0kT_e$ can then be calculated, where $k$ is Boltzmann's constant and assuming $T_e$=7000~K and a 100\% ionization fraction. Finally $\beta_{th}$ can be determined for each data point.

\subsection{Background pulsar search}

In Section \ref{probes} we explained how measurements of the RM and DM of pulsars can directly yield $B_{||}$ along a line of sight.  In order to identify possible pulsars intersecting our sample of \HII\ regions we carried out a search of the ATNF pulsar database\footnote{On-line catalogue at {\tt http://www.atnf.csiro.au/research/pulsar/psrcat} v.1.38, November 2009}  \citep{mhth05}. Of our sample of five \HII\ regions, Sh 2-27 has three pulsars and Sh 2-264 has one pulsar that overlaps the \HII\ region. DM-derived distances from the pulsar catalog are unreliable, as they rely on models of electron density in the Galaxy \citep{tc93,cl02} that do not include the \HII\ regions in our sample. In order to ensure accurate distance estimates, thus selecting only pulsars that lie behind \HII\ regions, we selected only the pulsars with distances derived from measurements of annual parallax.  This left a single pulsar, PSR J1643-1224 \citep{lnl+95}, which lies behind Sh 2-27.  We perform a consistency check on our results for Sh 2-27 in Section \ref{results}, using the measured properties of this pulsar. 

\section{Results}
\label{results}


Figure \ref{figure1} shows the observed \halpha\ surface brightness (grayscale) and rotation measures (circles) for each of our \HII\ regions, displayed in a stereographic projection. Blue and red circles indicate negative and positive RMs, respectively. The position of PSR J1643-1224 behind Sh 2-27 is denoted by a filled green circle.  \halpha\ maps were smoothed to a resolution of 1$^{\circ}$ (the resolution of the WHAM beam). For each background polarized source we measured the I$_{H\alpha}$ of the \halpha\ pixel closest to the NVSS source. In the case of two overlapping pixels the source was assigned the mean I$_{H\alpha}$ of the two. 

It appears that the magnitude of the RM of polarized sources is greater inside the \HII\ region boundary than outside the boundary, particularly for Sh 2-27 and Sh 2-264 with the highest \halpha\ surface brightnesses. In order to inspect this relationship more clearly, we made plots of RM vs. I$_{H\alpha}$ for each region. Figure \ref{figure2} shows these plots of the observed RM vs. I$_{H\alpha}$ for polarized background sources in each region of sky shown in Figure \ref{figure1}. Black circles indicate NVSS sources that lie outside the boundary of the \HII\ region.  Red and blue circles denote sources from \citet{tss09} with positive and negative RMs, respectively, that lie behind the \HII\ region. 

The plots in Figure~\ref{figure2} show unambiguously that, for polarized sources whose radiation has passed through an \HII\ region, there is a positive correlation between I$_{H\alpha}$ and $|$RM$|$. This is most clearly evident for Sh 2-27 and Sh 2-264, which have a large number of background polarized radio sources. The fact that RM has a consistent sign (positive or negative) within each region indicates that the magnetic field is largely coherent. Such a large-scale coherent magnetic field in the ionised gas of \HII\ regions has been predicted by the MHD simulations of \citet{arthur+11}. In addition to this coherent signal there is a scatter in RM $\sim \pm$ 100 rad~m$^{-2}$, which is likely related to the random component of the magnetic field. We defer our discussion of random magnetic fields to a future paper. 

For Sh 2-220 (the California Nebula), inspection of Figure \ref{figure1}(e) initially suggests that the \HII\ region is introducing a negative RM to the extragalactic sources. However, in Figure \ref{figure2}(e) we also see four positive RMs between 60$-$90 rad~m$^{-2}$ that apparently lie within the boundary of the \HII\ region.  Closer investigation reveals that these points lie on the northern edge of the Sh 2-220 \HII\ region, which has a very sharp drop-off in \halpha\ surface brightness due to the interaction between the \HII\ region and a molecular cloud \citep{ee78}.  The diameter of the WHAM beam is 1$^{\circ}$ and any polarized source along the edge of the \HII\ region is assigned a mean value of I$_{H\alpha}$ for the whole WHAM beam. We therefore assume that the \halpha\ surface brightness plotted on Figure \ref{figure1} for these four sources is artificially high due to beam averaging effects. Future, high-resolution \halpha\ observations could confirm this interpretation.

Next, we present the calculations of electron density, $n_0$  and the magnetic field parallel to the line-of-sight, $B_{||}$. The measured value of I$_{H\alpha}$ includes emission from the \HII\ region plus contributions from the foreground and background warm ionized material in the Galaxy.  In order to account for this, we calculated the mean \halpha\ surface brightness from a representative region of sky surrounding the \HII\ region boundary and subtracted this value from all \halpha\ surface brightness measurements within the boundary of the \HII\ region. Similarly, the extrinsic RM (or in some cases, a systematic RM gradient across the field) was subtracted from all RM values within the \HII\ region boundary. Details of the I$_{H\alpha}$ boundary conditions and magnitudes of the extrinsic RM and I$_{H\alpha}$ corrections in the \HII\ regions are given in Table \ref{table2}. 

After correcting the measured values of RM and I$_{H\alpha}$ for foreground and background contributions, we converted I$_{H\alpha}$ to EM using Equation~(\ref{eq1}), assuming that $T_e =$ 7000~K \citep{rey88,mrh06} and adopting values of E$_{B-V}$ from the literature (listed in Table \ref{table1}). We then calculated the electron density, $n_0$, of the \HII\ region along the sightline to each background polarized source. We defined a central position for each \HII\ region and calculated the path-length through the region for each background source, using the equation $L = 2[R^2 - D^2  sin^2(\theta)]^{1/2}$  where $R$ is the radius of the sphere, $D$ is the distance from the observer to the center of the sphere and $\theta$ is the angle between the line that goes from the observer to the center and the line that goes from the observer to the background source in question. The electron density $n_0$ derived from each \halpha\ measurements was then calculated using Equation (\ref{eq2}). We adopted a volume filling factor $f$ = 0.1, which is representative of the filling factors estimated for a number of \HII\ regions of this size in the literature \citep{kwe+89, hbg+82}. Finally, we calculated $B_{||}$ for the \HII\ region towards each polarized source using Equation (\ref{eq3}). 

Figure \ref{figure3} shows a plot of log $n_0$ against log $B_{||}$ for our 5 \HII\ regions. The black line represents equilibrium between magnetic and thermal pressures, which is discussed in Section \ref{pressure}. There is a significant scatter of both $n_0$ and $B_{||}$ within each region. This scatter in $B_{||}$ is possibly related to turbulence, as the magnitude of $B_{||}$ changes with the direction of the magnetic field vector with respect to the line of sight.  Some of this scatter may also be due to the uncertainty in $E_{B-V}$ along individual lines of sight, which affects the calculation of $B_{||}$. Another contributing factor to the scatter in the data may be the uncertainty in the electron filling factor for each sightline. In Figure~\ref{figure3} (bottom right) we have indicated with an arrow the direction and distance by which each data point will move if the filling factor changes from $f=0.1$ (plotted) to values of $f=0.04$ and $f=0.25$. This range in $f$ is representative of filling factors measured in \HII\ regions. The standard deviation of the optical depth, $\sigma_{\tau}$, for each \HII\ region is displayed as an arrow on the left-hand side of Figure~\ref{figure3}. Colored arrows (left) indicate the distance and direction each data point would move were the optical depth reduced by one standard deviation. 

Table \ref{table3} lists the derived values of the diameter, $\overline{B_{||}}$ and $\overline{n_0}$ of each \HII\ region, where bars indicate median values for each \HII\ region.  The estimated median electron densities for each region (approximately equal to the density of hydrogen atoms, $n(H)$, assuming a fully ionized medium) have values 1.5 $< n_0 <$ 14.1 cm$^{-3}$.  The line-of-sight magnetic field strengths in our \HII\ regions range from $-$6.3 to $+$2.2 $\mu$G, similar in magnitude to estimates of the magnetic field strength in the diffuse ISM using other methods \citep{hc05,crutcher07}. This suggests that the magnetic field in these large-diameter \HII\ regions may be the ambient magnetic field of the Galaxy that is `lit up' by virtue of the high electron density of the \HII\ region.  Estimates of the uncertainties in $B_{||}$ and $n_0$ are quoted, although the derived magnetic field strengths may be over- or underestimated by approximately a factor of three, due to possible correlation of fluctuations in electron density and magnetic fields in turbulent media \citep{bss+03}.

Following our analysis, we carried out a consistency check on our results by using the RM and DM of PSR J1643-1224 to independently measure the filling factor and magnetic field strength of Sh 2-27. Pulsar J1643-1224 was discovered by \citet{lnl+95} and lies at a distance (derived by annual parallax) of 455 $\pm$ 144 ~pc \citep{vbc+09}. The RM of PSR J1643-1224 is $-$263$\pm$ 15 rad~m$^{-2}$ \citep{mh04} and its DM is 62.4 cm$^{-3}$ pc \citep{tsb+99}. We removed the foreground contributions to the RM and DM due to the magnetoionic medium of the Galaxy in the direction of PSR J1643-1224 by subtracting the RM = $+$6.5$\pm$ 10 rad~m$^{-2}$ \citep{manchester74} and DM = 10.7 cm$^{-6}$ \citep{hlk+04} of PSR B1604-00 from the measured values for PSR J1643-1224.  We used PSR B1604-00 for this correction as it lies at a similar distance to PSR J1643-1224 (590 $\pm$~148 pc), derived by the electron density model of \citet{tc93} and has co-ordinates close to, but outside the boundary of Sh 2-27. 

Putting the foreground-corrected RM and DM into the relationship $B_{||} = \frac{RM}{0.81 \rm{DM}}$, we find that $B_{||} =  - 6.4~\mu$G. This is consistent with the value $B_{||}$=$-$6.1 $\pm$ 2.8 $\mu$G that we derived from polarized extragalactic sources, within the uncertainties. \halpha\ composite maps published by \citet{fink03} allowed us to calculate a residual I$_{H\alpha}$ (after subtracting the mean background of 0.9~R) of 70 R at the position of PSR J1643-1224.  Using Equation (\ref{eq1}), assuming a temperature of 7000~K and using the value of E$_{B-V}$ from \citet{sfd98} to derive the dust correction factor $e^{\tau} = 2.79$, we calculated the EM due to the ionized gas in Sh 2-27 towards PSR J1643-1224 to be EM = 395 cm$^{-6}$~pc. We find that $n_0 = \frac{EM}{DM} = 7.6$ cm$^{-3}$, which is broadly consistent with the median value 10.6 $\pm$ 2.8 cm$^{-3}$ derived from background RMs for Sh 2-27, within the estimated uncertainties. We also calculate a filling factor, $f = \frac{DM^2}{EM~L} = 0.2$ for Sh 2-27, which is comparable to the value $f$ = 0.1 that we have assumed in our calculations for density and magnetic field strength.

\section{Discussion}
\label{discussion}

\subsection{Comparison with previous results}
\label{comparison}
The strengths of magnetic fields in and around \HII\ regions have previously been investigated using a wide variety of observational methods, including Zeeman splitting of radio recombination lines \citep{th77,rgv+91,roshi07} and  {\sc Hi}, OH and H$_2$O maser lines \citep{brm92,crm+96,str01,bt01}. Most of these spectral lines originate in relatively dense regions in neutral and molecular gas, therefore none of these observational probes is suited to examining the $B$ vs. $n$ relation in warm ionized gas, where the results of existing observations are very poorly constrained.

Depolarization of radio synchrotron emission can be used to probe magnetic fields in \HII\ regions, provided the distance to the \HII\ region is known. \citet{gld+99} studied the Faraday rotation of Galactic synchrotron emission by the \HII\ region W4, placing an upper limit on the average magnetic field strength of $\langle B_{||}\rangle$ $<$ 20 $\mu$G, where angle brackets indicate a line-of-sight average. A similar approach was taken by \citet{gdm+01}, who modeled the depolarization of Galactic radio synchrotron emission by the \HII\ regions RCW 94 and ``void 1" (which we identify as RCW 105) with $n_e \sim 20$ cm$^{-3}$, to obtain an estimate of the random magnetic field strengths $B_r$ $\sim$5 $\mu$G (void 1) and $B_r$ = 1.2 $\mu$G (RCW 94). Using the fact that depolarization is negligible towards the edges of the voids were the gradient in RM is most severe, and assuming that the random and uniform components of the magnetic field are approximately equal, they determined that $B_{||} \sim$ 5 $\mu$G in void 1. For RCW 94, the polarization profile was modelled and the assumption that the uniform and random components of the field are equal was adopted, leading to the estimated value of $B_{||} \sim$ 1.2 $\mu$G. 

\citet{hc80} and \citet{hct81} were the first to use Faraday rotation of background polarized point-sources to probe the line-of-sight strength and orientation of the magnetic fields in \HII\ regions. They estimated the parallel component of the magnetic field strengths in 4 \HII\ regions, finding field strengths of the order $B_{||} \sim$1$-$10 $\mu$G. \citet{hct81} constrained the magnetic field strength of Sh 2-264 to the range $B_{||}$ = 0 to $+$2.6 $\mu$G. This agrees with the magnitude and direction for the magnetic field of $+$2.2 $\pm$ 1.6 $\mu$G derived by us, within the errors, although they used an average value of $n_e$ for the whole \HII\ region that was much lower than the median value of $n_0$ that we derived.  More recently, \citet{mwk+03} used the RM and DM of two pulsars to estimate the line-of-sight component of the uniform magnetic field in the \HII\ region Sh 2-205, finding that $B_{||} \sim $+5.7 $\mu$G. In each of these studies, only one or two polarized background sources was available to probe each region and the foreground correction was poorly constrained. In our study, we have used the Faraday rotation of a large sample of polarized extragalactic point sources to isolate $B_{||}$ in 5 \HII\ regions. As we have a far larger set of measurements than \citet{hc80,hct81} or \citet{mwk+03}, the value and scatter of RM in our \HII\ regions are much better characterised. As well as employing a large number of background sources in our study, we have also used detailed maps of \halpha\ surface brightness. We are therefore able to take many independent measurements of $n_e$ and $B_{||}$ for each \HII\ region.  In addition, the large number of extragalactic point sources in each field enable us to carry out a correction for the RM contribution of the background and foreground to the \HII\ region, thus isolating the magnetic field within a discrete volume of interstellar space.

A similar technique to ours was employed by \citet{wsi+09}, who described a `Faraday rotation anomaly' caused by an ionised shell surrounding the Cygnus OB1 association.  Within a 2$^{\circ}$--5$^\circ$ region surrounding the massive star cluster, they confirmed that the RMs of background radio sources change by several hundred rad~m$^{-2}$. The authors presented a model of an ionized plasma shell driven by stellar winds, in which the magnetic field of the interstellar medium is enhanced by compression. The RM distribution of the 9 polarized sources measured by Whiting et al. (2009) could not confirm or refute the possibility of an enhanced magnetic field in the shell, however the intensity distribution of \halpha\ emission from the WHAM survey was consistent with a limb-brightened shell surrounding Cygnus OB1. A similar scenario was proposed by \citet{stil+09}, who produced 3-dimensional MHD models of a magnetized superbubble and predicted the distribution of RMs and the gas density that would result from this bubble. They found that, if the line of sight is along the Galactic magnetic field, the largest rotation measures are associated with a thick shell at the rim of the \HII\ region.  Comparing this model to our data, the RM data from \citet{tss09} show no evidence for the existence of a shell with an enhanced magnetic field in any of our 5 \HII\ regions, although we cannot rule out a thin shell. To measure an enhanced magnetic field in a thin shell would require a far denser grid of background RMs. We searched for evidence \halpha\ emission from an ionized shell in our \HII\ regions, but in each case we found no systematic increase in \halpha\ surface brightness towards the rim. The only exception was Sh 2-220, where the \HII\ region is interacting with a molecular cloud, producing a sharp edge to the \halpha\ surface brightness. For these reasons, we do not consider a magnetized shell model to reflect the data in our sample of large-diameter \HII\ regions.

\subsection{The reliability of RMs from the Taylor et al. (2009) catalogue}

Mao et al. (2010) claimed to find discrepancies between the RMs from Taylor et al. (2009) produced by angle fitting between two adjacent spectral channels and their own RM measurements of the same sources, calculated using multi-channel RM synthesis. This difference was attributed by Mao et al. (2010) to incorrect RM assignments in the Taylor catalogue caused by multiple RM components that produce non-linearity in the polarization angle against $\lambda^2$ relation. Stil et al. (2011) addressed to these claims in Section 4.2 of their paper, reporting that the variance in the correlation of RMs common to the Taylor and Mao catalogues was underestimated compared to the errors quoted in each catalogue. Stil et al. (2011) also pointed out that in the Mao et al. (2010) RM sample, RMs with amplitude more than 20 rad m$^{-2}$ are much more likely to have a large error, which introduces a correlation between RM error and RM amplitude. This appears to explain the discrepancy between the two data sets, and also indicates that the uncertainties in RMs that satisfy $|$RM$|$ $<$ 8 rad~m$^{-2}$ are approximately 22$\%$ larger than quoted in the Taylor catalogue. If this interpretation is correct, it would have the effect of increasing the smallest error bars on our plots by 22$\%$ (Figure 2).  Given the small size of the error bars for these sources, such an increase would have a negligible effect on our analysis. An observational study by \citet{law+11} measured the RMs of 37 polarized sources using RM synthesis on data from the Allen Telescope Array. They compared their results with published RMs from Taylor et al. (2009) and found that in that catalogue, the RMs of sources with fluxes above 200 mJy do not suffer from n$\pi$ ambiguities and have at least a 95$\%$ reliability. \citet{vaneck+11} also compared their RMs to those of Taylor et al. (2009), finding a 96$\%$ correlation between the data sets.

\subsection{Magnetic vs. thermal pressure in \HII\ regions}
\label{pressure}

The magnetic field of the Galaxy is largely confined to the disk by the pressure of the gas threaded by the field (Parker 1966). Measuring the individual components of the total pressure in the ISM is important in testing multi-phase models of hydrostatic support in the Galactic disk, allowing simulations of the regulation of galactic star formation rates \citep{oml10} and determination of the scale height of galactic disks \citep{ko09}.  Using our measurements of magnetic field strength and electron density, we were able to estimate the magnetic and thermal pressure in five \HII\ regions that lie in the local Galactic disk.  


Figure \ref{figure3} shows the relationship $B_{||}$ vs. $n$, for a total of 93 measurements of polarized sources behind 5 \HII\ regions.  For each data point we calculated the thermal pressure, magnetic pressure and hence their ratio $\beta_{th}$.  The median $\overline{P}_{mag}$, $\overline{P}_{ther}$, $\overline{\beta}_{th}$ and the ranges that encompass 68$\%$ of the values of $\overline{\beta}_{th}$, centered on the median for each \HII\ region are listed in Table~\ref{table3}.  Note that quoted $\overline{\beta}_{th}$ = $\overline{(P_{ther}/P_{mag})}$ for all the data points in an \HII\ region, not the ratio $\overline{P}_{ther}$/$\overline{P}_{mag}$ for that region. The black line on Figure~\ref{figure4} indicates equipartition between thermal and magnetic pressure, $2n_0kT_e = \frac{B_{tot}^2}{8\pi}$, assuming a temperature of 7000~K and a 100\% ionization fraction.  

Our data reveal a wide range of values of $\beta_{th}$ in our 5 \HII\ regions, with median values 2.0 $\le \overline{\beta}_{th} <$ 22.4. In Sh 2-264 and Sh 2-171 $\overline{\beta}_{th} >>$ 1 within the estimated uncertainties and  in the other three HII regions, $\overline{\beta}_{th}$ $\sim$2.  Values of $\beta_{th} \ge$ 1 suggest that the \HII\ regions are still in a thermally-evolving/expanding phase, i.e. their morphology is dominated by thermal pressure and is not yet shaped by the magnetic field of the surrounding medium \citep{ksg07}.   In molecular clouds and the cold neutral medium,  typical values of $\beta_{th}$ are $\sim$ 0.04 and $\sim$0.3, respectively \citep{crutcher99, hc05}.  Improving the method by which measurements of electron density are made (e.g. by studying multiple radio recombination lines or measuring radio continuum emission) may help to reduce the uncertainties in $\beta_{th}$. Another way to achieve this is to improve the number density of background polarized radio sources, thus allowing a more sophisticated removal of the foreground RM and reducing the measurement error in $B_{||}$. A density of 100 polarized sources per square degree will likely be achieved by a planned survey of the RM sky using the Australian Square Kilometer Array Pathfinder \citep{glt10}.

The strengths of the magnetic fields in our \HII\ regions are similar to estimates of the magnetic field strength in the diffuse ISM. This suggests that the magnetic field that we measure in these regions is simply the magnetic field of the warm ionized medium of the Galaxy. The influence of the OB star at the center of the \HII\ region is simply to ionize the surrounding few tens of parsecs, allowing us to use the Faraday rotation of background sources to measure the properties of the Galactic magnetic field along the line of sight. In this way, we have measured the strength and orientation of the Galaxy's magnetic field in five discrete, isolated regions of the warm ionized medium.  Future work on a larger sample of \HII\ regions will assist in constraining $\beta_{th}$ in a wider range of \HII\ regions.  

\subsection{The magnetic field $-$ density relation}
In Section 1 we introduced the $B$ vs. $n$ relation and its relevance to molecular cloud formation and the large-scale structure of the ISM. We now measure the slope of $B$ vs. $n$ in the low-density ISM by adding our data to existing observations of magnetic fields in atomic and molecular gas. 

The relationship between magnetic field and density in neutral gas and molecular clouds is well constrained by a wealth of observations \citep[see, e.g.][]{cht03}. This relationship in the ISM is well described by $B$ $\propto n^{\kappa}$, where the exponent $\kappa$ has a value of approximately 0.4$-$0.6 in molecular clouds \citep[e.g.][]{crutcher99}\footnotemark[2]\footnotetext[2]{The relationship $B$ $\propto \sigma_{v,tot} n^{\kappa}$, where $\sigma_{v,tot}$ is the 1-dimensional velocity dispersion of the gas is also sometimes used \citep{basu00}.}. In low density gas traced by neutral hydrogen and the Faraday rotation and dispersion of pulsar signals, existing observations suggest that the strength of the magnetic field is invariant across three orders of magnitude in density \citep{th86,hc05,crutcher07}. However, the $B$ vs. $n$ relationship in the diffuse ISM is very poorly constrained by observations, (see Section \ref{intro}) with a sparsely sampled density range and many observations being upper limits. Our results from low-density \HII\ regions (where $n_0 \sim 0.8 - 30$ cm$^{-3}$), in addition to existing measurements of magnetic field strengths in the ISM allow us to parameterize the $B$ vs. $n$ relationship across this range of densities. 

Figure \ref{figure5} shows a plot of $B_{||}$ vs. $n$ in the density range 1 $-$ 10$^7$ cm$^{-3}$, using a compilation of published data from \citet{cwh+10} (green circles and blue squares) overplotted with our data from extragalactic point sources seen through \HII\ regions (red diamonds). The plotted density, $n$, is that of hydrogen atoms and molecules in neutral and molecular gas respectively and is the electron density $n_0$ in \HII\ region clumps. We have plotted $B_{||}$, the quantity directly measured both by Zeeman splitting and Faraday rotation. The error bars on our data arise from the uncertainties in measurements of RM and \halpha\ surface brightness. There are additional uncertainties in E$_{B-V} $ and $f$, which were shown in Figure \ref{figure3}. We employed the Levenberg-Marquardt iteration technique \citep{More78} to carry out separate parametric non-linear least squares fits to the data in the dense molecular gas and in the low-density (\HII\ + {\sc Hi}) gas. Using a function of the form $|$B$_{||}|$ = $A n^{\kappa}$ in the density range $n >$ 480 cm$^{-3}$ (i.e. for the neutral and ionized medium), our fit yielded $\kappa$ = 0.66 $\pm$ 0.05, which is consistent with the slope quoted by \citet{cwh+10} for the same data. In the density range $n <$ 480 cm$^{-3}$, populated by {\sc Hi} data and our new measurements of \HII\ regions, our fit yielded $\kappa$ = 0.11 $\pm$ 0.09. Quoted uncertainties represent a 68$\%$ confidence interval.

Our data have allowed us to probe magnetic fields in discrete volumes of ionized gas with densities in the range 0.8 $-$ 30 cm$^{-3}$. The value of the density exponent $\kappa$ that we derived for the neutral plus ionized gas is very close to, but sightly above zero.  This means that neither the atomic hydrogen in the cold neutral medium nor the ionized gas in clumps within our \HII\ regions is strongly coupled to the magnetic field.  It may be that there is no coupling ($\kappa$ = 0) and we have underestimated our uncertainties. It seems more likely that we have chosen our density cutoff slightly higher than the limit for self-gravitating clouds \citep[for example,][chose a cutoff point at $n$= 300 cm$^{-3}$]{cwh+10}. Our data have significantly expanded the density parameter space for which magnetic field strengths  are known. Further observations of ionized gas at even lower densities will help to constrain this slope further. Such studies would benefit from using a measure of electron density that does not suffer from extinction, such as radio recombination lines or free-free radio continuum emission.  The uncertainties in our measured values of $B_{||}$ were affected to a large degree by the dust extinction correction, e$^\tau$, as well as the filling factor, $f$. Although the uncertainties in $f$ are very difficult to remove, maps of radio continuum emission would allow a more accurate determination of the total electron density along each line of sight.

\subsection{Magnetic field orientation in the local Galaxy}

Several authors have mapped out the magnetic field of the Galaxy using measurements of pulsar RMs and DMs \citep{ls89,rl94,hml+06,njk+08,hz07} although the uncertainties in the distances to most pulsars, which depend on models of the Galactic electron density \citep{tc93,cl02}, place strong constraints on the accuracy of the derived magnetic field configuration. Studies of the Galactic magnetic field have also been made by measuring Faraday rotation of extragalactic point sources \citep{bhg+07,rps08,tss09,mgh+10,nk10}. Pulsar RMs probe the electron density-weighted line-of-sight magnetic field between the pulsar and earth, whereas individual extragalactic RMs probe the electron density-weighted average magnetic field throughout the entire Galaxy. Both of these techniques are powerful and complementary, provided that the distances to pulsars are known well and the number of extragalactic point-sources is sufficient to subtract out local fluctuations in $B_{||}$ and $n_e$. What they cannot do well is measure the configuration of the magnetic field in the solar neighbourhood. Here in \HII\ regions, we have measured $B_{||}$ in five volumes of the local Galaxy spread around three quadrants, thus providing a valuable addition to these existing data on the configuration of the local magnetic field within 1 kpc of the Sun. At the time of writing, there were also 25 pulsars with measured RMs that have distance estimates from parallax measurements within the local volume. Using these combined data, we have attempted to map out the magnetic field configuration in the solar neighbourhood.  

Figure \ref{figure5} shows a map of the local 1 kpc of the Galaxy, viewed from three orthogonal directions, onto which we have plotted our 5 \HII\ regions and all the pulsars with both measured RMs and with published distances measured by parallax.  Both above and below the Galactic plane, our measurements of $B_{||}$ are consistent with a preferred direction of $B$ towards $l \sim$ 90$^{\circ}$ within 1 kpc of the Sun.  This agrees with the conclusions of \citet{manchester72,manchester74,tn80,hq94,rl94,hml+06,bhg+07} and \citet{njk+08}, who found that the uniform magnetic field in the Galactic plane is oriented counterclockwise in Sagittarius-Carina spiral arm.  Our data do not allow us to comment on the proposed reversal in the Scutum-Crux arm, proposed by \citet{ls89,rl94,fss+01,bhg+07}. Previous studies of large pulsar datasets \citep{hq94,hmq99} have claimed to see a weak vertical field component in the local neighbourhood. A study of extragalactic point-source RMs found a preferred vertical magnetic field direction below the Galactic plane but no preferred direction above it \citep{mgh+10}. Our RMs (X-Z and Y-Z projections) show no evidence of a preferred uniform magnetic field direction orthogonal to the Galactic plane in the solar neighbourhood.

\section{Conclusions}

Using the RMs of an ensemble of background polarized sources and maps of surface brightness from published \halpha\ surveys, we have measured the line-of-sight component of the magnetic field strength in 5 Galactic \HII\ regions. Within each \HII\ region, the line-of-sight component of the magnetic field is coherent. The median strength of $B_{||}$ in each region lies between 2.2 and 6.3~$\mu$G and the estimated median electron density ranges from 1.5 to 14.1 cm$^{-3}$, where the electrons are assumed to reside in clumps that occupy 10$\%$ of each sightline through the \HII\ region. For one of the \HII\ regions, Sh 2-27, we also verified our estimate of $B_{||}$ and $n_0$ using the RM of a background pulsar. 

With such a large number of background polarized sources, we have been able to subtract the RM due to the Galaxy and determine an independent measure of electron density and $B_{||}$ for each sightline for which we had a background RM. Our measurements offer a significant improvement on existing published data (or upper limits) in the low-density ionized ISM. Future, similar studies of \HII\ regions with a larger range of densities will constrain the relationship between $B_{||}$ and $n_0$ further.

Our measurements of $B_{||}$ and $n_0$ allowed us to estimate the magnetic and thermal pressure in 5 Galactic \HII\ regions. In two regions the thermal pressure was dominant and in the other three, thermal and magnetic pressure were approximately equal within the relatively large uncertainties of the measurements. This suggests that the \HII\ regions are moving towards an advanced stage of their formative phase, where the dominance of thermal pressure is diminishing and the external magnetic field is able to take a role in shaping the regions.

The consistent sign (positive or negative) of the RM within each \HII\ region indicates that the magnetic field is largely coherent. As a consequence, we were able to measure the direction of $B_{||}$ in five regions of the local Galaxy. Both above and below the Galactic plane, the magnetic field is oriented roughly towards $l$=90$^{\circ}$. There is no preferred vertical direction of the Galactic magnetic field apparent from our data. Future studies involving measurements of $B_{||}$ in a larger number of \HII\ regions will extend our understanding of the magnetic field in the local Galaxy.

\begin{acknowledgements}
L.H-S. and B.M.G. acknowledge the support of the Australian Research Council through grants DP0986386 and FF0561298. GJM is supported by a Fellowship from the Research Office of the University of Sydney. The authors thank Dick Crutcher for generously providing data on magnetic field strengths and densities in atomic hydrogen and molecular clouds. This research made use of the NASA Astrophysics Data System (ADS). The National Radio Astronomy Observatory is a facility of the National Science Foundation operated under cooperative agreement by Associated Universities, Inc. The Virginia Tech Spectral-Line Survey (VTSS), the Southern H-Alpha Sky Survey Atlas (SHASSA), and the Wisconsin H-Alpha Mapper (WHAM) are all funded by the National Science Foundation. SHASSA observations were obtained at Cerro Tololo Inter-American Observatory, which is operated by the Association of Universities for Research in Astronomy, Inc., under cooperative agreement with the National Science Foundation.
\end{acknowledgements}

{\it Facilities:} \facility{VLA}, \facility{WHAM}, \facility{SHASSA}, \facility{VTSS}

\bibliographystyle{apj}
\bibliography{lisarefs}

\begin{thebibliography}{86}
\expandafter\ifx\csname natexlab\endcsname\relax\def\natexlab#1{#1}\fi

\bibitem[{{Arthur} {et~al.}(2011){Arthur}, {Henney}, {Mellema}, {De Colle}, \&
  {V{\'a}zquez-Semadeni}}]{arthur+11}
{Arthur}, S.~J., {Henney}, W.~J., {Mellema}, G., {De Colle}, F., \&
  {V{\'a}zquez-Semadeni}, E. 2011, Accepted to MNRAS, ArXiv:1101.5510

\bibitem[{{Bailes} {et~al.}(1990){Bailes}, {Manchester}, {Kesteven}, {Norris},
  \& {Reynolds}}]{bmk+90}
{Bailes}, M., {Manchester}, R.~N., {Kesteven}, M.~J., {Norris}, R.~P., \&
  {Reynolds}, J.~E. 1990, \nat, 343, 240

\bibitem[{{Basu}(2000)}]{basu00}
{Basu}, S. 2000, \apjl, 540, L103

\bibitem[{{Beck}(2001)}]{beck01}
{Beck}, R. 2001, \ssr, 99, 243

\bibitem[{{Beck} {et~al.}(2003){Beck}, {Shukurov}, {Sokoloff}, \&
  {Wielebinski}}]{bss+03}
{Beck}, R., {Shukurov}, A., {Sokoloff}, D., \& {Wielebinski}, R. 2003, \aap,
  411, 99

\bibitem[{{Blitz} {et~al.}(1982){Blitz}, {Fich}, \& {Stark}}]{bfs82}
{Blitz}, L., {Fich}, M., \& {Stark}, A.~A. 1982, \apjs, 49, 183

\bibitem[{{Bloemhof} {et~al.}(1992){Bloemhof}, {Reid}, \& {Moran}}]{brm92}
{Bloemhof}, E.~E., {Reid}, M.~J., \& {Moran}, J.~M. 1992, \apj, 397, 500

\bibitem[{{Brisken} {et~al.}(2002){Brisken}, {Benson}, {Goss}, \&
  {Thorsett}}]{bbgt02}
{Brisken}, W.~F., {Benson}, J.~M., {Goss}, W.~M., \& {Thorsett}, S.~E. 2002,
  \apj, 571, 906

\bibitem[{{Brisken} {et~al.}(2003){Brisken}, {Thorsett}, {Golden}, \&
  {Goss}}]{btgg03}
{Brisken}, W.~F., {Thorsett}, S.~E., {Golden}, A., \& {Goss}, W.~M. 2003,
  \apjl, 593, L89

\bibitem[{{Brogan} \& {Troland}(2001)}]{bt01}
{Brogan}, C.~L. \& {Troland}, T.~H. 2001, \apj, 560, 821

\bibitem[{{Brown} {et~al.}(2007){Brown}, {Haverkorn}, {Gaensler}, {Taylor},
  {Bizunok}, {McClure-Griffiths}, {Dickey}, \& {Green}}]{bhg+07}
{Brown}, J.~C., {Haverkorn}, M., {Gaensler}, B.~M., {Taylor}, A.~R., {Bizunok},
  N.~S., {McClure-Griffiths}, N.~M., {Dickey}, J.~M., \& {Green}, A.~J. 2007,
  \apj, 663, 258

\bibitem[{{Chatterjee} {et~al.}(2009){Chatterjee}, {Brisken}, {Vlemmings},
  {Goss}, {Lazio}, {Cordes}, {Thorsett}, {Fomalont}, {Lyne}, \&
  {Kramer}}]{cbv+09}
{Chatterjee}, S., {Brisken}, W.~F., {Vlemmings}, W.~H.~T., {Goss}, W.~M.,
  {Lazio}, T.~J.~W., {Cordes}, J.~M., {Thorsett}, S.~E., {Fomalont}, E.~B.,
  {Lyne}, A.~G., \& {Kramer}, M. 2009, \apj, 698, 250

\bibitem[{{Chatterjee} {et~al.}(2004){Chatterjee}, {Cordes}, {Vlemmings},
  {Arzoumanian}, {Goss}, \& {Lazio}}]{ccv+04}
{Chatterjee}, S., {Cordes}, J.~M., {Vlemmings}, W.~H.~T., {Arzoumanian}, Z.,
  {Goss}, W.~M., \& {Lazio}, T.~J.~W. 2004, \apj, 604, 339

\bibitem[{{Condon} {et~al.}(1998){Condon}, {Cotton}, {Greisen}, {Yin},
  {Perley}, {Taylor}, \& {Broderick}}]{ccg+98}
{Condon}, J.~J., {Cotton}, W.~D., {Greisen}, E.~W., {Yin}, Q.~F., {Perley},
  R.~A., {Taylor}, G.~B., \& {Broderick}, J.~J. 1998, \aj, 115, 1693

\bibitem[{{Cordes} \& {Lazio}(2002)}]{cl02}
{Cordes}, J.~M. \& {Lazio}, T.~J.~W. 2002, arXiv:astro-ph/0207156

\bibitem[{{Crutcher} {et~al.}(2003){Crutcher}, {Heiles}, \& {Troland}}]{cht03}
{Crutcher}, R., {Heiles}, C., \& {Troland}, T. 2003, in Lecture Notes in
  Physics, Berlin Springer Verlag, Vol. 614, Turbulence and Magnetic Fields in
  Astrophysics, ed. {E.~Falgarone \& T.~Passot}, 155--181

\bibitem[{{Crutcher}(1999)}]{crutcher99}
{Crutcher}, R.~M. 1999, \apj, 520, 706

\bibitem[{{Crutcher}(2007)}]{crutcher07}
{Crutcher}, R.~M. 2007, in IAU Symposium, Vol. 242, IAU Symposium, ed.
  {J.~M.~Chapman \& W.~A.~Baan}, 47--54

\bibitem[{{Crutcher} {et~al.}(1996){Crutcher}, {Roberts}, {Mehringer}, \&
  {Troland}}]{crm+96}
{Crutcher}, R.~M., {Roberts}, D.~A., {Mehringer}, D.~M., \& {Troland}, T.~H.
  1996, \apjl, 462, L79

\bibitem[{{Crutcher} {et~al.}(2010){Crutcher}, {Wandelt}, {Heiles},
  {Falgarone}, \& {Troland}}]{cwh+10}
{Crutcher}, R.~M., {Wandelt}, B., {Heiles}, C., {Falgarone}, E., \& {Troland},
  T.~H. 2010, ApJ, 725, 466

\bibitem[{{Deller} {et~al.}(2009){Deller}, {Tingay}, {Bailes}, \&
  {Reynolds}}]{dtbr09}
{Deller}, A.~T., {Tingay}, S.~J., {Bailes}, M., \& {Reynolds}, J.~E. 2009,
  \apj, 701, 1243

\bibitem[{{Dennison} {et~al.}(1998){Dennison}, {Simonetti}, \&
  {Topasna}}]{dst98}
{Dennison}, B., {Simonetti}, J.~H., \& {Topasna}, G.~A. 1998, PASA, 15, 147

\bibitem[{{Dodson} {et~al.}(2003){Dodson}, {Legge}, {Reynolds}, \&
  {McCulloch}}]{dlrm03}
{Dodson}, R., {Legge}, D., {Reynolds}, J.~E., \& {McCulloch}, P.~M. 2003, \apj,
  596, 1137

\bibitem[{{Elmegreen} \& {Elmegreen}(1978)}]{ee78}
{Elmegreen}, D.~M. \& {Elmegreen}, B.~G. 1978, \apj, 219, 105

\bibitem[{{Fiedler} \& {Mouschovias}(1993)}]{fm93}
{Fiedler}, R.~A. \& {Mouschovias}, T.~C. 1993, \apj, 415, 680

\bibitem[{{Finkbeiner}(2003)}]{fink03}
{Finkbeiner}, D.~P. 2003, \apjs, 146, 407

\bibitem[{{Frick} {et~al.}(2001){Frick}, {Stepanov}, {Shukurov}, \&
  {Sokoloff}}]{fss+01}
{Frick}, P., {Stepanov}, R., {Shukurov}, A., \& {Sokoloff}, D. 2001, \mnras,
  325, 649

\bibitem[{{Gaensler} {et~al.}(2001){Gaensler}, {Dickey}, {McClure-Griffiths},
  {Green}, {Wieringa}, \& {Haynes}}]{gdm+01}
{Gaensler}, B.~M., {Dickey}, J.~M., {McClure-Griffiths}, N.~M., {Green}, A.~J.,
  {Wieringa}, M.~H., \& {Haynes}, R.~F. 2001, \apj, 549, 959

\bibitem[{{Gaensler} {et~al.}(2010){Gaensler}, {Landecker}, {Taylor}, \&
  {POSSUM Collaboration}}]{glt10}
{Gaensler}, B.~M., {Landecker}, T.~L., {Taylor}, A.~R., \& {POSSUM
  Collaboration}. 2010, in BAAS, Vol.~42, 515

\bibitem[{{Gaustad} {et~al.}(2001){Gaustad}, {McCullough}, {Rosing}, \& {Van
  Buren}}]{gmr+01}
{Gaustad}, J.~E., {McCullough}, P.~R., {Rosing}, W., \& {Van Buren}, D. 2001,
  \pasp, 113, 1326

\bibitem[{{Gray} {et~al.}(1999){Gray}, {Landecker}, {Dewdney}, {Taylor},
  {Willis}, \& {Normandeau}}]{gld+99}
{Gray}, A.~D., {Landecker}, T.~L., {Dewdney}, P.~E., {Taylor}, A.~R., {Willis},
  A.~G., \& {Normandeau}, M. 1999, \apj, 514, 221

\bibitem[{{Gwinn} {et~al.}(1986){Gwinn}, {Taylor}, {Weisberg}, \&
  {Rawley}}]{gtwr86}
{Gwinn}, C.~R., {Taylor}, J.~H., {Weisberg}, J.~M., \& {Rawley}, L.~A. 1986,
  \aj, 91, 338

\bibitem[{{Haffner} {et~al.}(2003){Haffner}, {Reynolds}, {Tufte}, {Madsen},
  {Jaehnig}, \& {Percival}}]{hrt+03}
{Haffner}, L.~M., {Reynolds}, R.~J., {Tufte}, S.~L., {Madsen}, G.~J.,
  {Jaehnig}, K.~P., \& {Percival}, J.~W. 2003, \apjs, 149, 405

\bibitem[{{Han} {et~al.}(2006){Han}, {Manchester}, {Lyne}, {Qiao}, \& {van
  Straten}}]{hml+06}
{Han}, J.~L., {Manchester}, R.~N., {Lyne}, A.~G., {Qiao}, G.~J., \& {van
  Straten}, W. 2006, \apj, 642, 868

\bibitem[{{Han} {et~al.}(1999){Han}, {Manchester}, \& {Qiao}}]{hmq99}
{Han}, J.~L., {Manchester}, R.~N., \& {Qiao}, G.~J. 1999, \mnras, 306, 371

\bibitem[{{Han} \& {Qiao}(1994)}]{hq94}
{Han}, J.~L. \& {Qiao}, G.~J. 1994, \aap, 288, 759

\bibitem[{{Han} \& {Zhang}(2007)}]{hz07}
{Han}, J.~L. \& {Zhang}, J.~S. 2007, \aap, 464, 609

\bibitem[{{Heiles} \& {Chu}(1980)}]{hc80}
{Heiles}, C. \& {Chu}, Y. 1980, \apjl, 235, L105

\bibitem[{{Heiles} {et~al.}(1981){Heiles}, {Chu}, \& {Troland}}]{hct81}
{Heiles}, C., {Chu}, Y., \& {Troland}, T.~H. 1981, \apjl, 247, L77

\bibitem[{{Heiles} \& {Crutcher}(2005)}]{hc05}
{Heiles}, C. \& {Crutcher}, R. 2005, in Lecture Notes in Physics, Berlin
  Springer Verlag, Vol. 664, Cosmic Magnetic Fields, ed. {R.~Wielebinski \&
  R.~Beck}, 137

\bibitem[{{Heitsch} {et~al.}(2007){Heitsch}, {Slyz}, {Devriendt}, {Hartmann},
  \& {Burkert}}]{heitsch+07}
{Heitsch}, F., {Slyz}, A.~D., {Devriendt}, J.~E.~G., {Hartmann}, L.~W., \&
  {Burkert}, A. 2007, \apj, 665, 445

\bibitem[{{Heitsch} {et~al.}(2009){Heitsch}, {Stone}, \& {Hartmann}}]{hsh09}
{Heitsch}, F., {Stone}, J.~M., \& {Hartmann}, L.~W. 2009, \apj, 695, 248

\bibitem[{{Herter} {et~al.}(1982){Herter}, {Briotta}, {Gull}, {Shure}, \&
  {Houck}}]{hbg+82}
{Herter}, T., {Briotta}, Jr., D.~A., {Gull}, G.~E., {Shure}, M.~A., \& {Houck},
  J.~R. 1982, \apj, 262, 164

\bibitem[{{Hobbs} {et~al.}(2004){Hobbs}, {Lyne}, {Kramer}, {Martin}, \&
  {Jordan}}]{hlk+04}
{Hobbs}, G., {Lyne}, A.~G., {Kramer}, M., {Martin}, C.~E., \& {Jordan}, C.
  2004, \mnras, 353, 1311

\bibitem[{{Kaspi} {et~al.}(1994){Kaspi}, {Taylor}, \& {Ryba}}]{ktr94}
{Kaspi}, V.~M., {Taylor}, J.~H., \& {Ryba}, M.~F. 1994, \apj, 428, 713

\bibitem[{{Kassim} {et~al.}(1989){Kassim}, {Weiler}, {Erickson}, \&
  {Wilson}}]{kwe+89}
{Kassim}, N.~E., {Weiler}, K.~W., {Erickson}, W.~C., \& {Wilson}, T.~L. 1989,
  \apj, 338, 152

\bibitem[{{Kim} \& {Ostriker}(2006)}]{kimo06}
{Kim}, W. \& {Ostriker}, E.~C. 2006, \apj, 646, 213

\bibitem[{{Koyama} \& {Ostriker}(2009)}]{ko09}
{Koyama}, H. \& {Ostriker}, E.~C. 2009, \apj, 693, 1346

\bibitem[{{Krumholz} {et~al.}(2007){Krumholz}, {Stone}, \& {Gardiner}}]{ksg07}
{Krumholz}, M.~R., {Stone}, J.~M., \& {Gardiner}, T.~A. 2007, \apj, 671, 518

\bibitem[{{Law} {et~al.}(2011){Law}, {Gaensler}, {Bower}, {Backer},
  {Bauermeister}, {Croft}, {Forster}, {Gutierrez-Kraybill}, {Harvey-Smith},
  {Heiles}, {Hull}, {Keating}, {MacMahon}, {Whysong}, {Williams}, \&
  {Wright}}]{law+11}
{Law}, C.~J., {Gaensler}, B.~M., {Bower}, G.~C., {Backer}, D.~C.,
  {Bauermeister}, A., {Croft}, S., {Forster}, R., {Gutierrez-Kraybill}, C.,
  {Harvey-Smith}, L., {Heiles}, C., {Hull}, C., {Keating}, G., {MacMahon}, D.,
  {Whysong}, D., {Williams}, P.~K.~G., \& {Wright}, M. 2011, \apj, 728, 57

\bibitem[{{Lorimer} {et~al.}(1995){Lorimer}, {Nicastro}, {Lyne}, {Bailes},
  {Manchester}, {Johnston}, {Bell}, {D'Amico}, \& {Harrison}}]{lnl+95}
{Lorimer}, D.~R., {Nicastro}, L., {Lyne}, A.~G., {Bailes}, M., {Manchester},
  R.~N., {Johnston}, S., {Bell}, J.~F., {D'Amico}, N., \& {Harrison}, P.~A.
  1995, \apj, 439, 933

\bibitem[{{Lyne} \& {Smith}(1989)}]{ls89}
{Lyne}, A.~G. \& {Smith}, F.~G. 1989, \mnras, 237, 533

\bibitem[{{Madsen} {et~al.}(2006){Madsen}, {Reynolds}, \& {Haffner}}]{mrh06}
{Madsen}, G.~J., {Reynolds}, R.~J., \& {Haffner}, L.~M. 2006, \apj, 652, 401

\bibitem[{{Manchester}(1972)}]{manchester72}
{Manchester}, R.~N. 1972, \apj, 172, 43

\bibitem[{{Manchester}(1974)}]{manchester74}
---. 1974, \apj, 188, 637

\bibitem[{{Manchester} \& {Han}(2004)}]{mh04}
{Manchester}, R.~N. \& {Han}, J.~L. 2004, \apj, 609, 354

\bibitem[{{Manchester} {et~al.}(2005){Manchester}, {Hobbs}, {Teoh}, \&
  {Hobbs}}]{mhth05}
{Manchester}, R.~N., {Hobbs}, G.~B., {Teoh}, A., \& {Hobbs}, M. 2005, 129, 1993

\bibitem[{{Mao} {et~al.}(2010){Mao}, {Gaensler}, {Haverkorn}, {Zweibel},
  {Madsen}, {McClure-Griffiths}, {Shukurov}, \& {Kronberg}}]{mgh+10}
{Mao}, S.~A., {Gaensler}, B.~M., {Haverkorn}, M., {Zweibel}, E.~G., {Madsen},
  G.~J., {McClure-Griffiths}, N.~M., {Shukurov}, A., \& {Kronberg}, P.~P. 2010,
  \apj, 714, 1170

\bibitem[{{Mitra} {et~al.}(2003){Mitra}, {Wielebinski}, {Kramer}, \&
  {Jessner}}]{mwk+03}
{Mitra}, D., {Wielebinski}, R., {Kramer}, M., \& {Jessner}, A. 2003, \aap, 398,
  993

\bibitem[{{More}(1978)}]{More78}
{More}, J.~J. 1978, in Numerical Analysis, Springer-Verlag: Berlin, Vol. 630,
  Lecture Notes in Mathematics, ed. {G.A. Watson}, 105--116

\bibitem[{{Nota} \& {Katgert}(2010)}]{nk10}
{Nota}, T. \& {Katgert}, P. 2010, \aap, 513, A65

\bibitem[{{Noutsos} {et~al.}(2008){Noutsos}, {Johnston}, {Kramer}, \&
  {Karastergiou}}]{njk+08}
{Noutsos}, A., {Johnston}, S., {Kramer}, M., \& {Karastergiou}, A. 2008,
  \mnras, 386, 1881

\bibitem[{{Ostriker} {et~al.}(2010){Ostriker}, {McKee}, \& {Leroy}}]{oml10}
{Ostriker}, E.~C., {McKee}, C.~F., \& {Leroy}, A.~K. 2010, \apj, 721, 975

\bibitem[{{Padoan} \& {Nordlund}(1999)}]{pn99}
{Padoan}, P. \& {Nordlund}, {\AA}. 1999, \apj, 526, 279

\bibitem[{{Piontek} \& {Ostriker}(2007)}]{po07}
{Piontek}, R.~A. \& {Ostriker}, E.~C. 2007, \apj, 663, 183

\bibitem[{{Rand} \& {Lyne}(1994)}]{rl94}
{Rand}, R.~J. \& {Lyne}, A.~G. 1994, \mnras, 268, 497

\bibitem[{{Reynolds}(1988)}]{rey88}
{Reynolds}, R.~J. 1988, \apj, 333, 341

\bibitem[{{Roberts} {et~al.}(1991){Roberts}, {Goss}, {van Gorkom}, \&
  {Leahy}}]{rgv+91}
{Roberts}, D.~A., {Goss}, W.~M., {van Gorkom}, J.~H., \& {Leahy}, J.~P. 1991,
  \apjl, 366, L15

\bibitem[{{Roshi}(2007)}]{roshi07}
{Roshi}, D.~A. 2007, \apjl, 658, L41

\bibitem[{{Roy} {et~al.}(2008){Roy}, {Pramesh Rao}, \& {Subrahmanyan}}]{rps08}
{Roy}, S., {Pramesh Rao}, A., \& {Subrahmanyan}, R. 2008, \aap, 478, 435

\bibitem[{{Sarma} {et~al.}(2001){Sarma}, {Troland}, \& {Romney}}]{str01}
{Sarma}, A.~P., {Troland}, T.~H., \& {Romney}, J.~D. 2001, \apjl, 554, L217

\bibitem[{{Schlegel} {et~al.}(1998){Schlegel}, {Finkbeiner}, \&
  {Davis}}]{sfd98}
{Schlegel}, D.~J., {Finkbeiner}, D.~P., \& {Davis}, M. 1998, \apj, 500, 525

\bibitem[{{Shull} \& {van Steenberg}(1985)}]{ss85}
{Shull}, J.~M. \& {van Steenberg}, M.~E. 1985, \apj, 294, 599

\bibitem[{{Sota} {et~al.}(2008){Sota}, {Ma{\'{\i}}z-Apell{\'a}niz}, {Walborn},
  \& {Shida}}]{smw+08}
{Sota}, A., {Ma{\'{\i}}z-Apell{\'a}niz}, J., {Walborn}, N.~R., \& {Shida},
  R.~Y. 2008, in Revista Mexicana de Astronomia y Astrofisica Conference
  Series, Vol.~33, 56--56

\bibitem[{{Stil} {et~al.}(2009){Stil}, {Wityk}, {Ouyed}, \& {Taylor}}]{stil+09}
{Stil}, J., {Wityk}, N., {Ouyed}, R., \& {Taylor}, A.~R. 2009, \apj, 701, 330

\bibitem[{{Taylor} {et~al.}(2009){Taylor}, {Stil}, \& {Sunstrum}}]{tss09}
{Taylor}, A.~R., {Stil}, J.~M., \& {Sunstrum}, C. 2009, \apj, 702, 1230

\bibitem[{{Taylor} \& {Cordes}(1993)}]{tc93}
{Taylor}, J.~H. \& {Cordes}, J.~M. 1993, \apj, 411, 674

\bibitem[{{Thomson} \& {Nelson}(1980)}]{tn80}
{Thomson}, R.~C. \& {Nelson}, A.~H. 1980, \mnras, 191, 863

\bibitem[{{Toscano} {et~al.}(1999){Toscano}, {Britton}, {Manchester}, {Bailes},
  {Sandhu}, {Kulkarni}, \& {Anderson}}]{tsb+99}
{Toscano}, M., {Britton}, M.~C., {Manchester}, R.~N., {Bailes}, M., {Sandhu},
  J.~S., {Kulkarni}, S.~R., \& {Anderson}, S.~B. 1999, \apjl, 523, L171

\bibitem[{{Troland} \& {Heiles}(1977)}]{th77}
{Troland}, T.~H. \& {Heiles}, C. 1977, \apj, 214, 703

\bibitem[{{Troland} \& {Heiles}(1986)}]{th86}
---. 1986, \apj, 301, 339

\bibitem[{{Vall{\'e}e}(1995)}]{vallee95}
{Vall{\'e}e}, J.~P. 1995, \apss, 234, 1

\bibitem[{{Van Eck} {et~al.}(2011){Van Eck}, {Brown}, {Stil}, {Rae}, {Mao},
  {Gaensler}, {Shukurov}, {Taylor}, {Haverkorn}, {Kronberg}, \&
  {McClure-Griffiths}}]{vaneck+11}
{Van Eck}, C.~L., {Brown}, J.~C., {Stil}, J.~M., {Rae}, K., {Mao}, S.~A.,
  {Gaensler}, B.~M., {Shukurov}, A., {Taylor}, A.~R., {Haverkorn}, M.,
  {Kronberg}, P.~P., \& {McClure-Griffiths}, N.~M. 2011, \apj, 728, 97

\bibitem[{{Verbiest} {et~al.}(2009){Verbiest}, {Bailes}, {Coles}, {Hobbs}, {van
  Straten}, {Champion}, {Jenet}, {Manchester}, {Bhat}, {Sarkissian}, {Yardley},
  {Burke-Spolaor}, {Hotan}, \& {You}}]{vbc+09}
{Verbiest}, J.~P.~W., {Bailes}, M., {Coles}, W.~A., {Hobbs}, G.~B., {van
  Straten}, W., {Champion}, D.~J., {Jenet}, F.~A., {Manchester}, R.~N., {Bhat},
  N.~D.~R., {Sarkissian}, J.~M., {Yardley}, D., {Burke-Spolaor}, S., {Hotan},
  A.~W., \& {You}, X.~P. 2009, \mnras, 400, 951

\bibitem[{{Whiting} {et~al.}(2009){Whiting}, {Spangler}, {Ingleby}, \&
  {Haffner}}]{wsi+09}
{Whiting}, C.~A., {Spangler}, S.~R., {Ingleby}, L.~D., \& {Haffner}, L.~M.
  2009, \apj, 694, 1452

\bibitem[{{Zweibel}(2002)}]{zweibel02}
{Zweibel}, E.~G. 2002, \apj, 567, 962

\end{thebibliography}

\clearpage

\begin{deluxetable}{crlrccccc}
\tabletypesize{\scriptsize}
\rotate
\tablewidth{0pt}
\tablewidth{0pt}
\tablehead{
\colhead{\HII\ region} & \colhead{$l$} &  \colhead{$b$} &  \colhead{d} &  \colhead{Ionizing Star} &   \colhead{Spectral Type} & \colhead{z} & \colhead{N}  &   \colhead{E$_{B-V}$} 
\\
 &  &  &  \colhead{(pc)} & &  &  \colhead{(pc)}& &
}
\tablecaption{\scriptsize{Adopted parameters for the sample of 5 \HII\ regions studied in this work. The Galactic co-ordinates $l,b$ define the central position of the \HII\ region. The estimated distance, d, to each \HII\ region is listed in parsecs. We include the name of the principle ionizing star near the center of each \HII\ region, along with its spectral type. The vertical distance from the Galactic plane, $z$ is given in parsecs. The number of useful polarized sources, N, from the catalog of \protect\citet{tss09} behind each \HII\ region is also given. Column 9 shows the color excess: E$_{B-V}$ is the mean value from \protect\citet{sfd98} within the \HII\ region boundary defined in Table~2.}} 
\startdata
Sh 2-27   & 8.0            & $+$23.5      &   146\tablenotemark{a}    & \zetaoph\       & O9.0~V     & $+$59& 57    &  0.47  \\
Sh 2-264 & 195.0 & $-$12     &  386\tablenotemark{a}     & \lambdaori\          & O8~III        & $-$80  & 17    &  0.45    \\
Sivan 3    & 144.5    & $+$14      &   1010\tablenotemark{b} & \alphacam\        & O9.5~Ia    & $+$244 & 10    &  0.32  \\
Sh 2-171 & 118.0       & $+$6     &   840\tablenotemark{c}   & BD+66 1673  & O5~V        &  $- $81  & 5      &  1.40   \\
Sh 2-220 & 161.3    & $-$12.5    &   398\tablenotemark{b}   & \chiper\              & O7.5~III     &  $-$89  & 4      &  0.28   \\
\enddata
\label{table1}
\tablenotetext{a}{\citet{smw+08}}
\tablenotetext{b}{\citet{ss85}}
\tablenotetext{c}{\citet{bfs82}}
\end{deluxetable}

\begin{deluxetable}{cccccll}
\tabletypesize{\tiny}
\tablewidth{0pt}
\tablehead{
\colhead{\HII\ region} & \colhead{On Region} & \colhead{I$_{H\alpha}$ On} &  \colhead{Off Region}  & \colhead{Extrinsic I$_{H\alpha}$} &  \colhead{Extrinsic RM} &  \colhead{Survey} 
\\
 & \colhead{Boundary}  & \colhead{(R)} & \colhead{Boundary} & \colhead{(R)} & \colhead{rad m$^{-2}$} &
}
\tablecaption{\scriptsize{Boundary conditions and details of extrinsic RM and EM subtraction for the \HII\ regions.  Pixels that lie within an `On Region Boundary' (column 2) in Galactic co-ordinates that also have I$_{H\alpha}$ greater than `I$_{H\alpha}$ On' (defined in column 3) are defined as being part of the \HII\ region. Pixels that lie within the `Off Region Boundary' (column 4) and have I$_{H\alpha}$ less than `I$_{H\alpha}$ On' were used to subtract the Galactic background and foreground emission that does not originate from the \HII\ region. The `Extrinsic I$_{H\alpha}$' is the mean \halpha\ surface brightness in the off region. Where an extrinsic RM gradient is quoted, $l$ is the galactic longitude, in degrees.}} 
\startdata
Sh 2-27  & 0$^{\circ}$  $<$ $l$ $<$ +15$^{\circ}$     &    20 & $-$5$^{\circ}$ $<$ $l$ $<$ $+$15$^{\circ}$     & 4.7   & $+$2.2 $l$ - 14.4 & SHASSA \\
         & +15$^{\circ}$ $<$ $b$ $<$ +35$^{\circ}$          &         &  +15$^{\circ}$ $<$ $b$ $<$ +35$^{\circ}$              &       &                         &        \\
\\
Sh 2-264 & $r$ $<$ 4$^{\circ}\!$.5  & 40 & 4$^{\circ}\!$.5 $< r <$ 6$^{\circ}$  & 17.4  & $+$18.8                    & SHASSA \\
         & $l,b$: $+$164$^{\circ}\!$.5,$-$12$^{\circ}$     &       & $l,b$: $+$164$^{\circ}\!$.5,$-$12$^{\circ}$                     &       &                         &        \\
\\
Sivan 3  & $r < $4$^{\circ}\!$.5 & 10 & 4$^{\circ}\!$.5 $< r <$ 5$^{\circ}\!.5$ 10  & 4.5   & $-$41.6             & WHAM   \\
         & $l,b$: $+$145$^{\circ}$,+14$^{\circ}$         &       &  $l,b$: +145$^{\circ}$,+14$^{\circ}$                        &       &                         & \& VTSS\\
\\
Sh 2-171 & +116$^{\circ}$  $<$ $l$ $<$ +125$^{\circ}$      & 7  & +120$^{\circ}$  $<$ $l$ $<$ +125$^{\circ}$           & 5.2   & $-$58.4                 & WHAM   \\
         & +4$^{\circ}$  $<$ $b$ $<$ +8$^{\circ}$          &      &  0$^{\circ}$  $<$ $b$ $<$ +10$^{\circ}$                         &       &                         &        \\
\\
Sh 2-220 &  +155$^{\circ}$  $<$ $l$ $<$ +165$^{\circ}$   & 20 & +155$^{\circ}$  $<$ $l$ $<$ +170$^{\circ}$              & 9.2   & $+$5.2 $l$ - 799  & WHAM   \\
         & $-$15$^{\circ}$ $<$ $b$ $<$ $-$10$^{\circ}$   &    &    $-$15$^{\circ}$ $<$ $b$ $<$ $-$10$^{\circ}$             &       &                         &        \\
\enddata
\label{table2}
\end{deluxetable}

\begin{deluxetable}{cccccccccc}
\tabletypesize{\scriptsize}
\rotate
\tablewidth{0pt}
\tablehead{
\colhead{HII region} & \colhead{Diameter} &  \colhead{$\overline{n}_0$}       &  \colhead{${\sigma}_{\overline{n}_0}$} &   \colhead{$\overline{B}_{||}$} &   \colhead{${\sigma}_{\overline{B}_{||}}$}   &  \colhead{$\overline{P}_{mag}$} & \colhead{$\overline{P}_{ther}$} & \colhead{$\overline{\beta}_{th}$} & \colhead{${\sigma}_{\overline{\beta}_{th}}$}
\\
                                     &  \colhead{(pc)}              &  \colhead{(cm$^{-3}$)}  &  \colhead{(cm$^{-3}$)}           &   \colhead{($\mu$G)} &   \colhead{($\mu$G)} &  \colhead{(dyn~cm$^{-2}$)} & \colhead{(dyn~cm$^{-2}$)}   &  &    
                     }
\tablecaption{\scriptsize{Derived values from the study of RMs of polarized sources behind 5 \HII\ regions. The diameter of each \HII\ region is quoted in parsecs, assuming each region is spherical. $\overline{n}_0$ is the median density of free electrons in the clumps within an \HII\ region and $\overline{B}_{||}$ is the median parallel component of the magnetic field strength. $\overline{B}_{||}$ and ${\sigma}_{\overline{n}_0}$ are ranges of these values that encompass 68\% of the measurements, centered on the median. $P_{mag}$ and $P_{ther}$ were calculated for each data point and here we quote the median of all values, $\overline{P}_{mag}$ and $\overline{P}_{ther}$ for each \HII\ region. The ratio of thermal to magnetic pressure was calculated for each data point. The median value, $\overline{\beta}_{th}$, and the 68\% range, ${\sigma}_{\overline{\beta}_{th}}$ are quoted here.}}
\startdata
Sh 2-27   & 34 & 10.6 & 2.8 &  $-$6.1  & 2.8 & 8.8$\times$10$^{-12}$ & 2.1$\times$10$^{-11}$& 2.2 & 2.3\\
Sh 2-264 & 60 & 9.7   & 2.0 &  $+$2.2 & 1.6 & 1.2$\times$10$^{-12}$ & 1.9$\times$10$^{-11}$ & 16.6 & 18.0\\
Sivan 3    & 166 & 1.5   & 0.3 &  $-$2.5  & 1.5 & 2.8$\times$10$^{-12}$ & 3.0$\times$10$^{-12}$ & 2.0 & 1.5\\
Sh 2-171 & 58 & 14.1 & 8.3 & $-$2.3  & 1.5 & 1.2$\times$10$^{-12}$ & 2.8$\times$10$^{-11}$ & 22.4 & 17.3\\
Sh 2-220 & 34 & 11.2   & 3.8 &  $-$6.3  & 2.4 &  1.7$\times$10$^{-11}$ & 2.2$\times$10$^{-11}$ & 2.3 & 1.7\\
\hline
\enddata
\label{table3}
\end{deluxetable}

\begin{figure}
\begin{center}
\figurenum{1}
\includegraphics[scale=0.6]{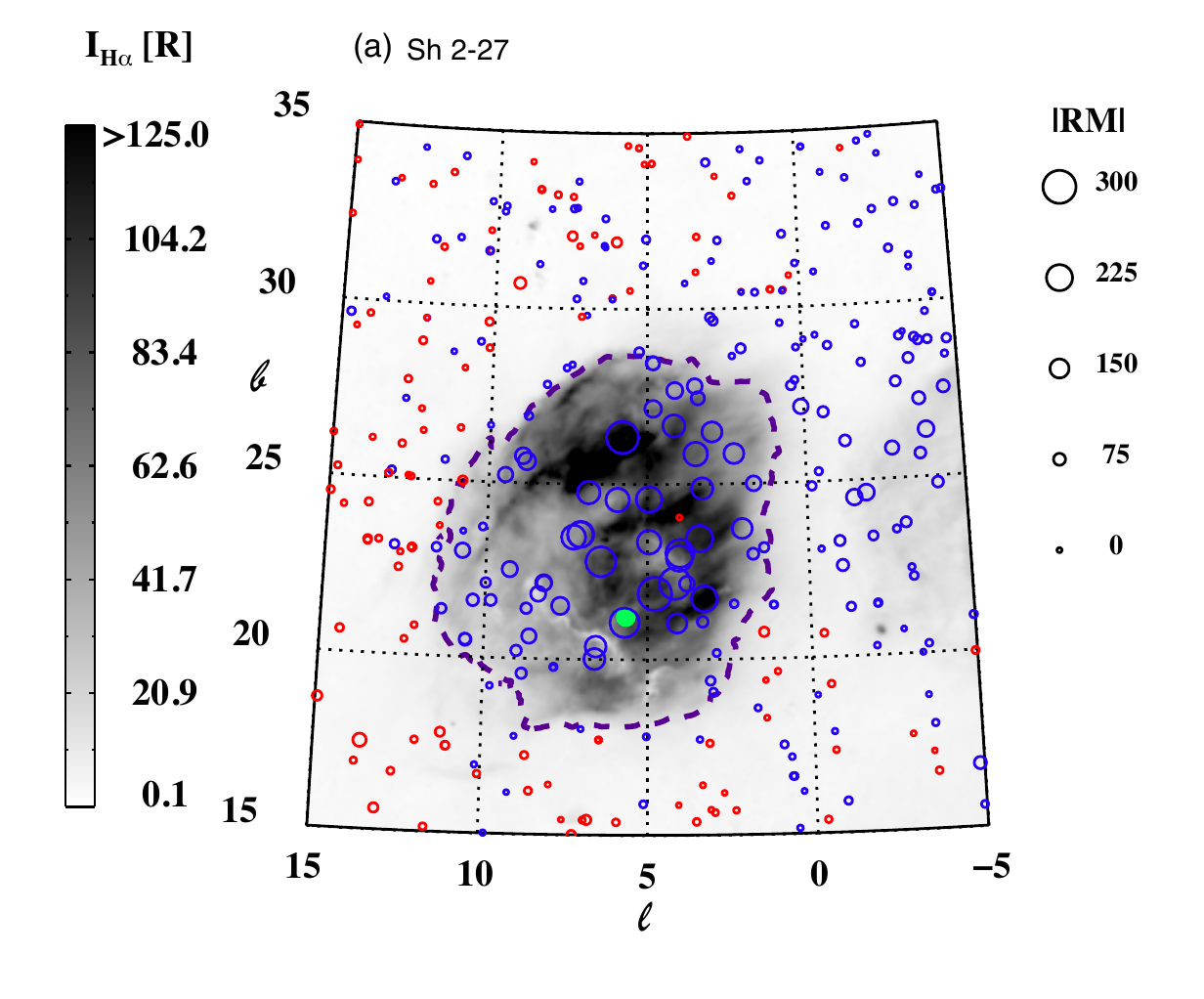}
\vspace{1cm}
\caption{Rotation measures (circles) and smoothed \halpha\ surface brightness with no extinction correction applied (grayscale) in (a) Sh 2-27, (b) Sh 2-264, (c) Sivan 3, (d) Sh 2-171, \& (e) Sh 2-220. Maps are labelled in Galactic co-ordinates. Blue and red circles indicate negative and positive RMs, respectively and the diameter of the circles are proportional to $|$RM$|$. The legend gives 5 examples of circle sizes and the corresponding $|$RM$|$ values in rad~m$^{-2}$. The green circle denotes the position of PSR J1643-1224. The dashed line denotes the boundary of the \HII\ region, as defined in Table \ref{table2}. The grayscale varies linearly between given minimum and maximum values and the colorbar shows the observed I$_{H\alpha}$ in rayleighs. Note that the dashed line indicating the boundary of each region is a guide, as the H$\alpha$ images have been smoothed.} \label{figure1}
\end{center}
\end{figure}

\begin{figure}
\begin{center}
\figurenum{1}
\includegraphics[scale=0.6]{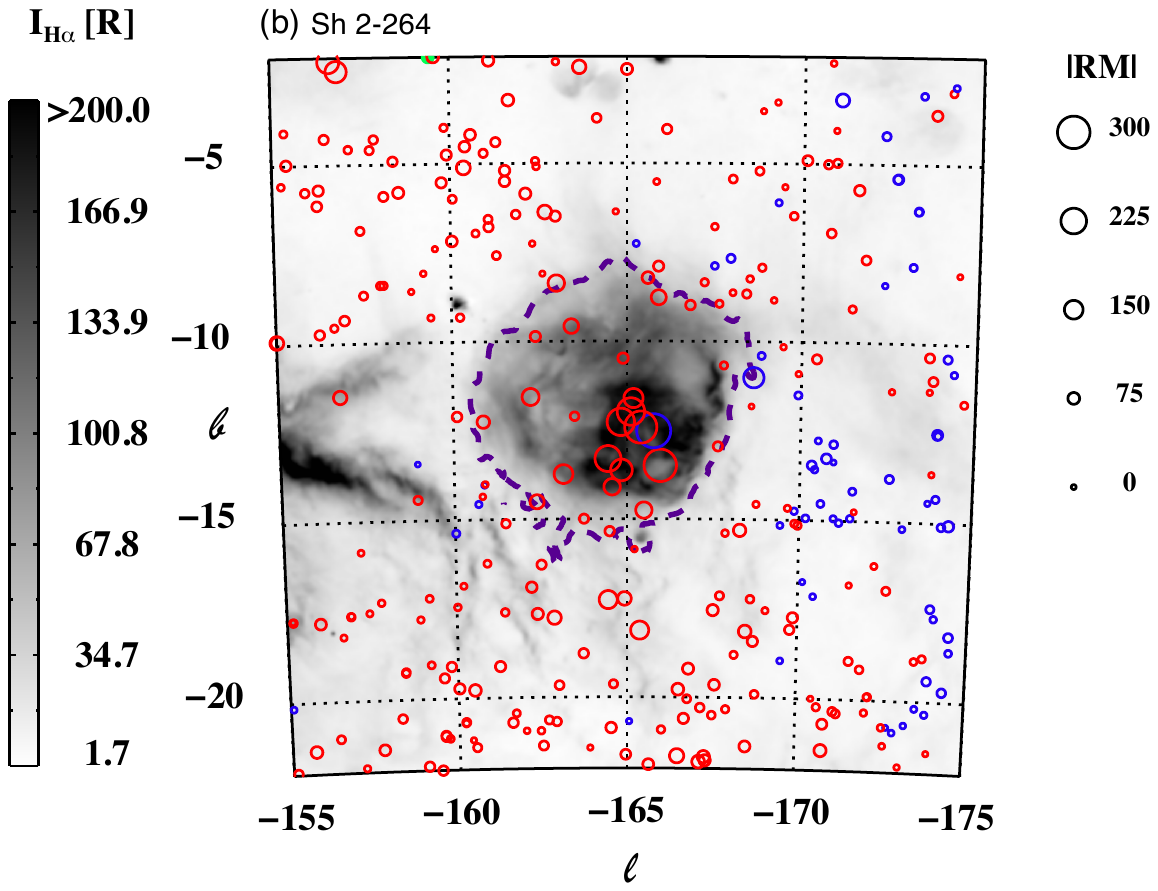}
\vspace{1cm}
\caption{continued (b) Sh 2-264} \label{figure1}
\end{center}
\end{figure}

\begin{figure}
\begin{center}
\figurenum{1}
\includegraphics[scale=0.6]{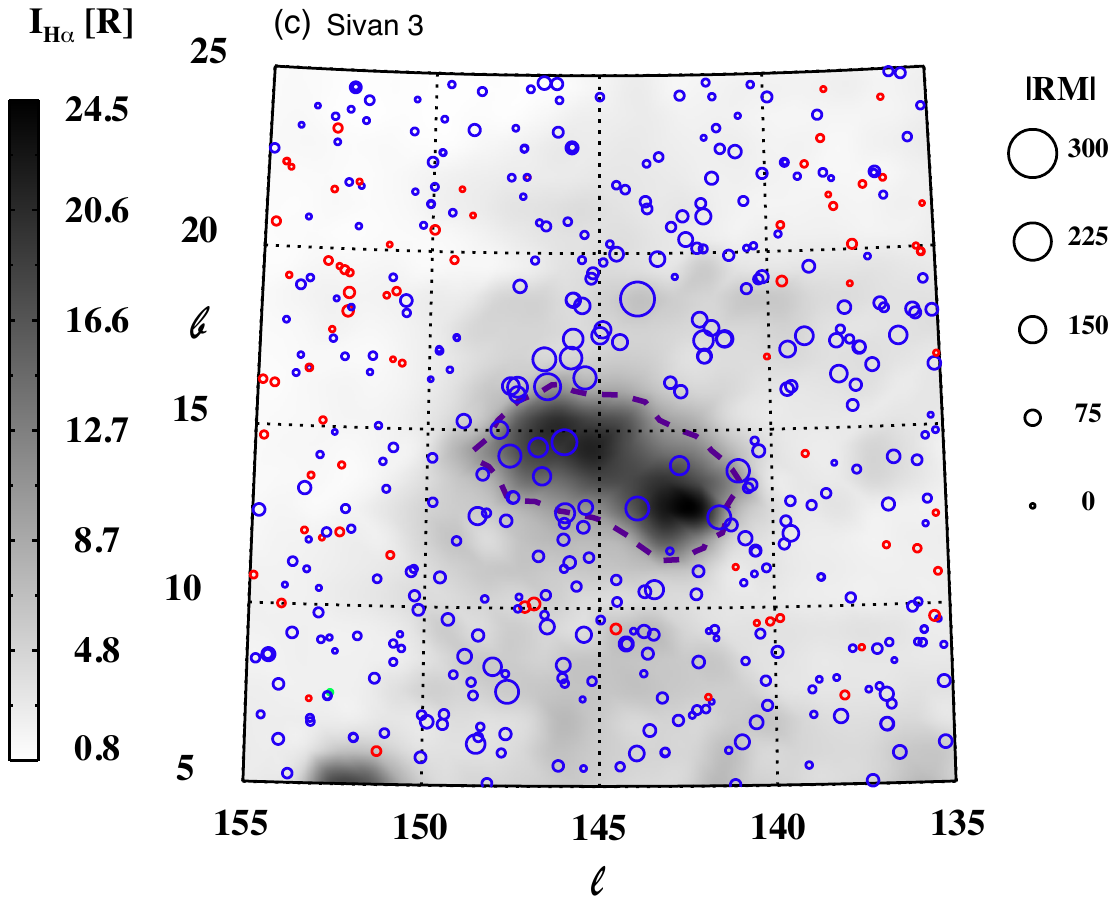}
\vspace{1cm}
\caption{continued (c) Sivan 3} \label{figure1}
\end{center}
\end{figure}

\begin{figure}
\begin{center}
\figurenum{1}
\includegraphics[scale=0.6]{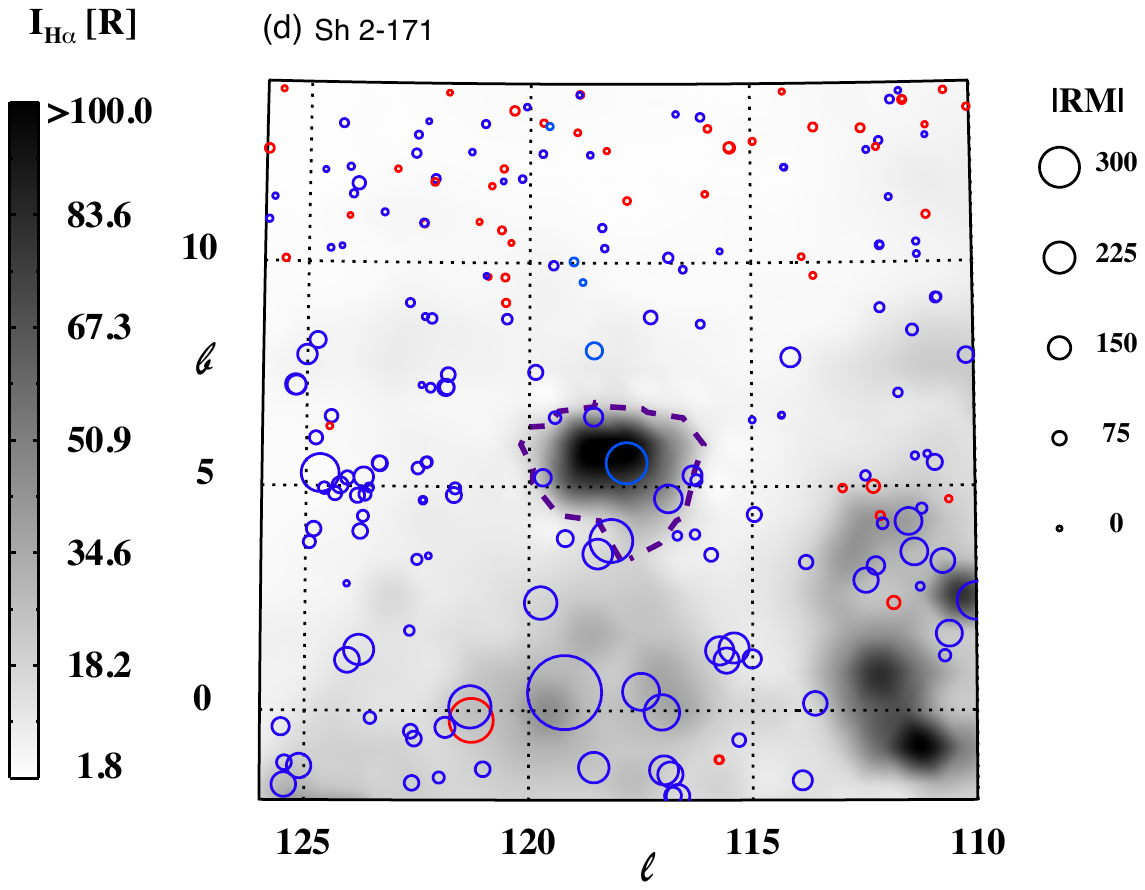}
\vspace{1cm}
\caption{continued (d) Sh 2-171} \label{figure1}
\end{center}
\end{figure}

\begin{figure}
\begin{center}
\figurenum{1}
\includegraphics[scale=0.6]{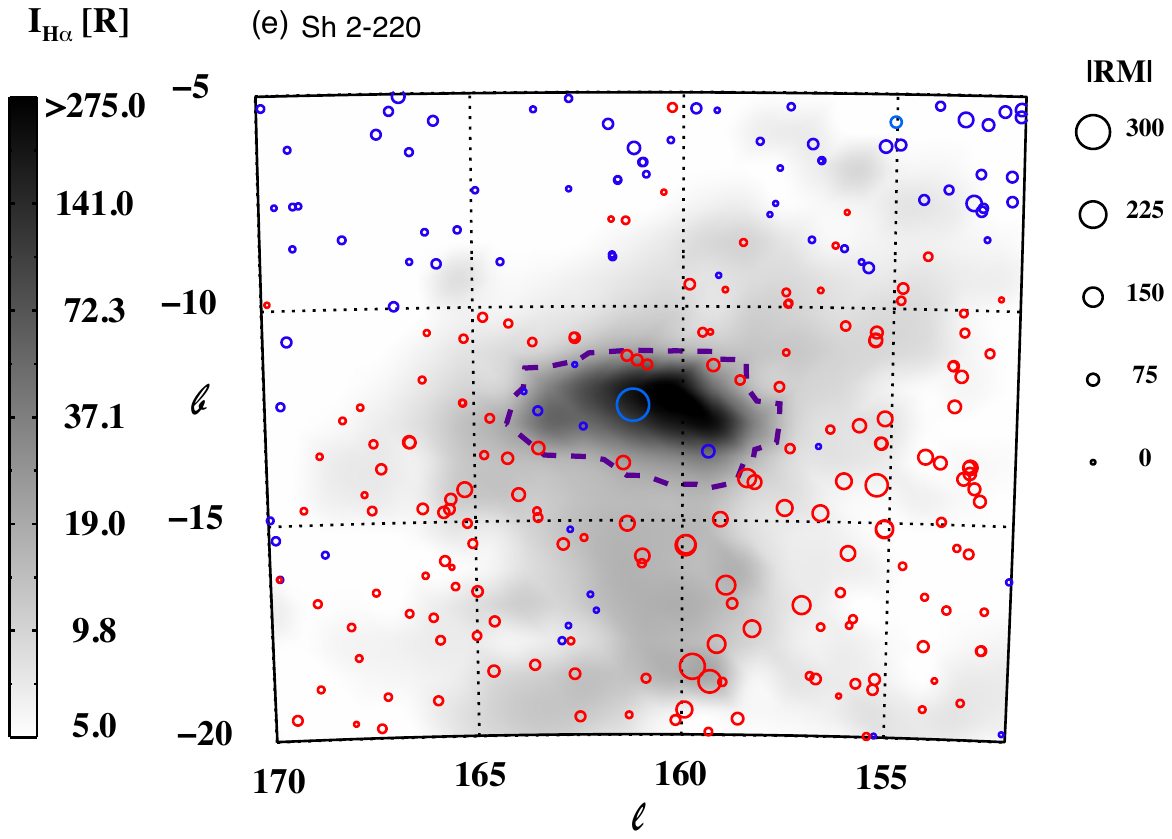}
\vspace{1cm}
\caption{continued (e) Sh 2-220} \label{figure1}
\end{center}
\end{figure}

\begin{figure}
\begin{center}
\figurenum{2}
\includegraphics[scale=0.7]{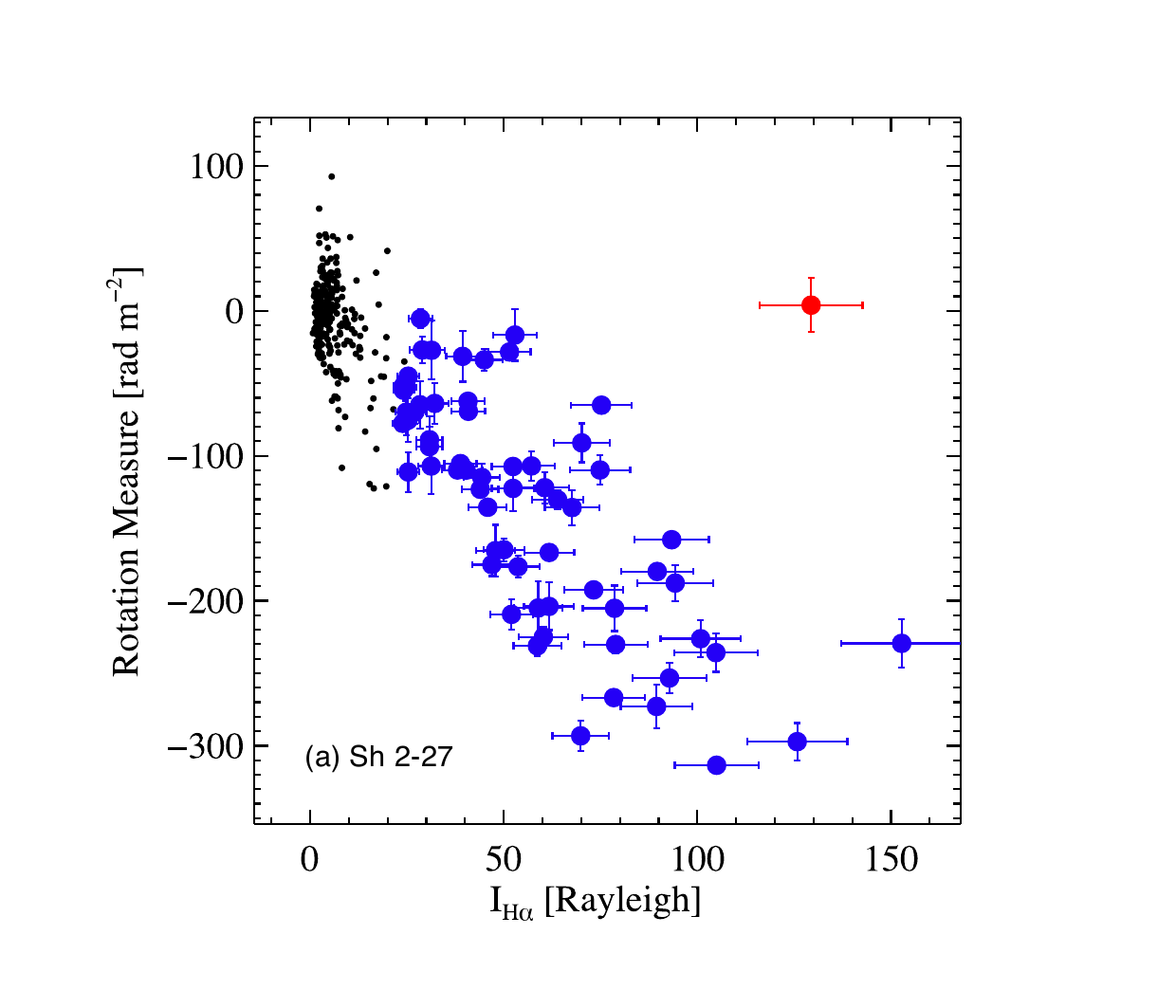}
\caption{I$_{H\alpha}$ vs. RM for (a) Sh 2-27, (b) Sh 2-264, (c) Sivan 3, (d) Sh 2-171 \& (e) Sh 2-220, showing the presence of coherent magnetic fields. Black points: extragalactic sources intersecting the sky plane outside the boundary of the \HII\ region, red and blue circles: point sources that intersect the \HII\ region and that have positive and negative RMs, respectively.}\label{figure2} 
\end{center}
\end{figure}

\begin{figure}
\begin{center}
\figurenum{2}
\includegraphics[scale=0.7]{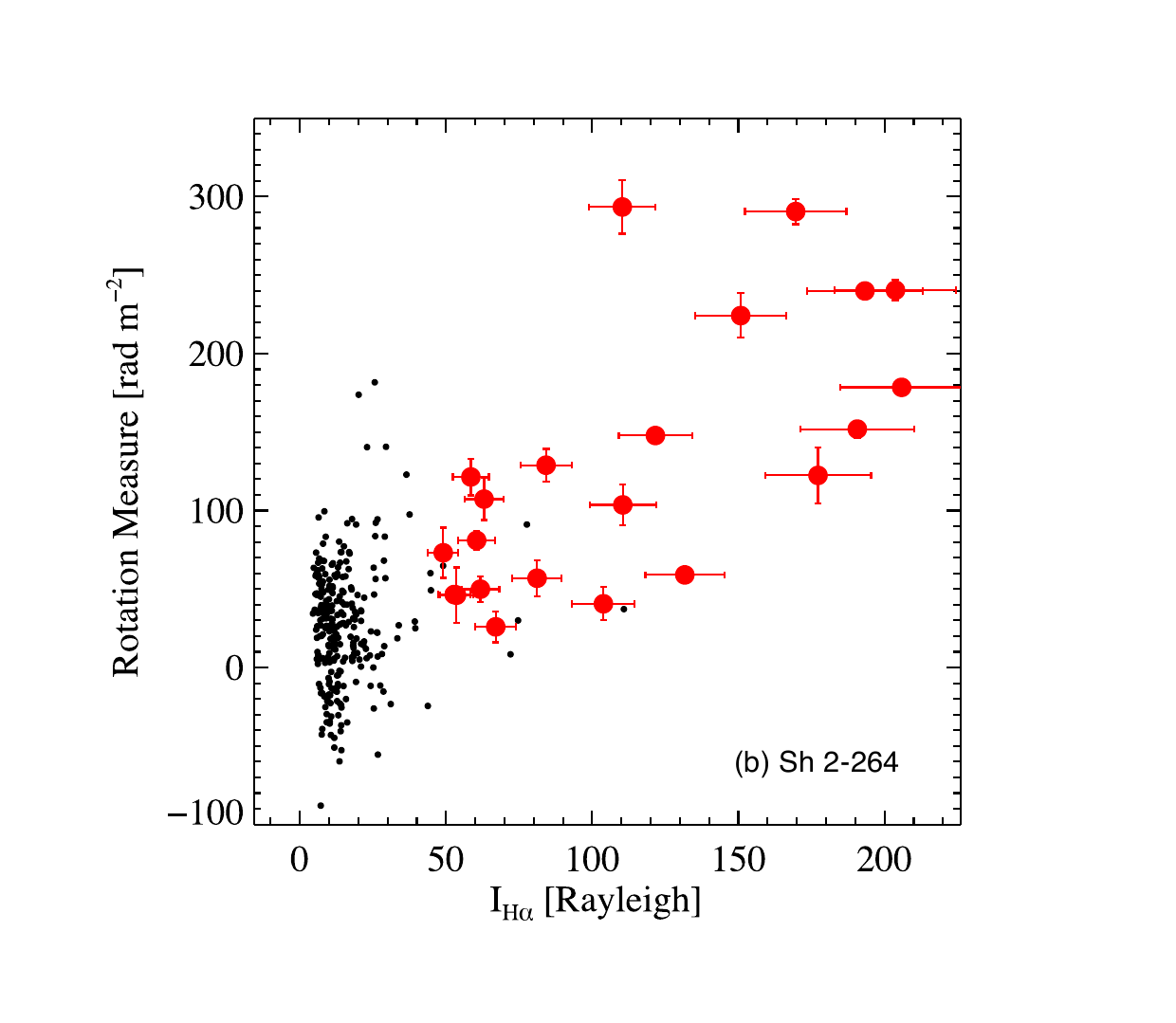}
\caption{continued, (b) Sh 2-264}\label{figure2} 
\end{center}
\end{figure}

\begin{figure}
\begin{center}
\figurenum{2}
\includegraphics[scale=0.7]{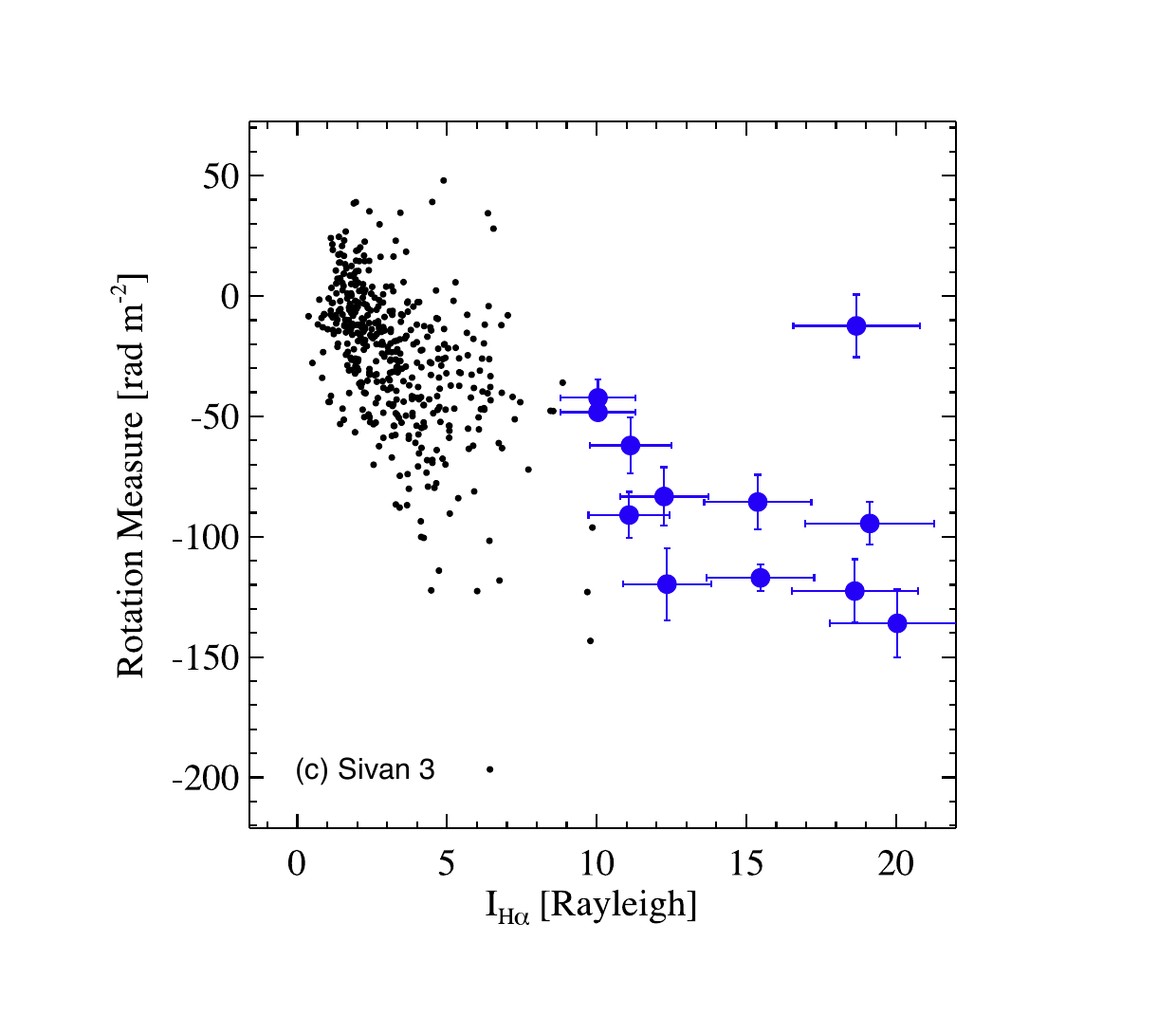}
\caption{continued, (c) Sivan 3}\label{figure2} 
\end{center}
\end{figure}

\begin{figure}
\begin{center}
\figurenum{2}
\includegraphics[scale=0.7]{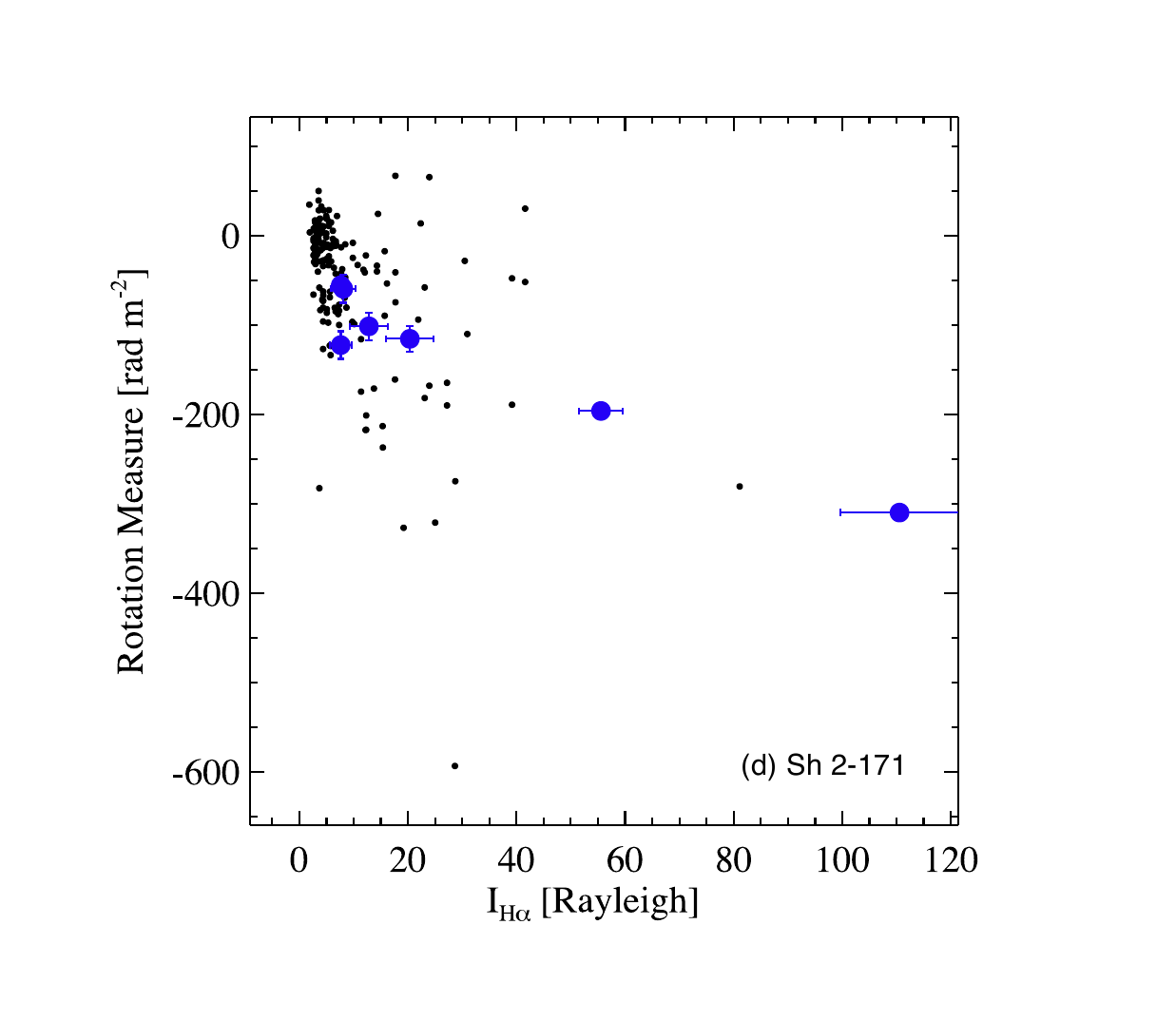}
\caption{continued, (d) Sh 2-171}\label{figure2} 
\end{center}
\end{figure}

\begin{figure}
\begin{center}
\figurenum{2}
\includegraphics[scale=0.7]{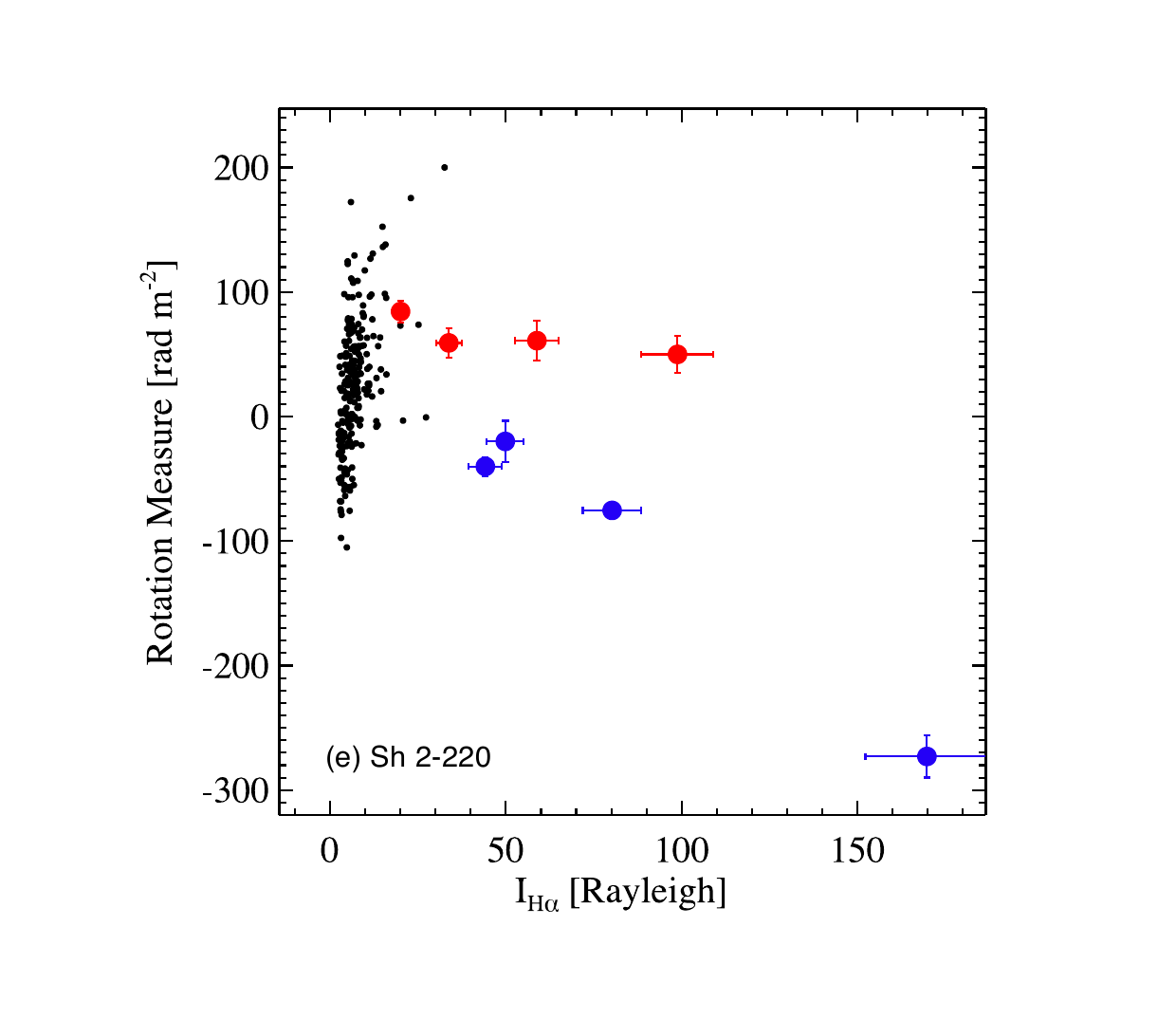}
\caption{continued, (e) Sh 2-220}\label{figure2} 
\end{center}
\end{figure}

\begin{figure}
\begin{center}
\figurenum{3}
\includegraphics[scale=0.6]{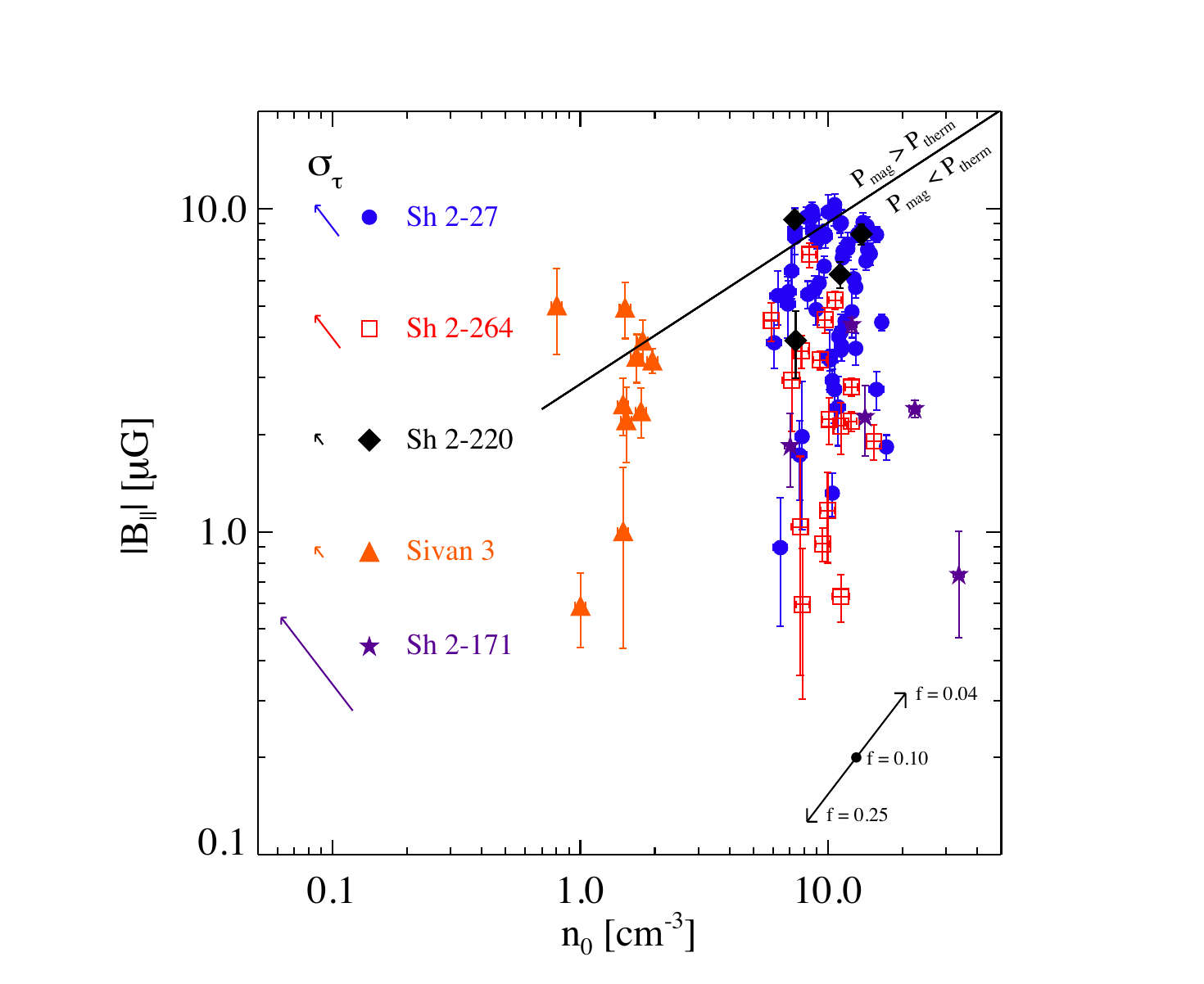}
\vspace{3cm}
\caption{\small{Parallel component of the magnetic field strength against electron density for sightlines through 5 Galactic \HII\ regions. Data have been corrected using the assumption that the dust contribution measured by \protect\citet{sfd98} lies completely in front of each \HII\ region. Colored arrows (left) indicate the displacement of any single data point if the \halpha\ optical depth were reduced by a factor equal to the standard deviation of the optical depth across that \HII\ region. Data are plotted assuming a filling factor of $f=0.1$. The black arrow (bottom right) indicates the movement of each data point resulting from a change in the electron filling factor, $f$, by a factor of 2.5 in either direction. The solid black line represents equilibrium between the magnetic and thermal pressure, assuming $T$=7000~K.} }
 \label{figure3}
\end{center}
\end{figure}

\begin{figure}
\begin{center}
\figurenum{4}
\includegraphics[scale=0.7]{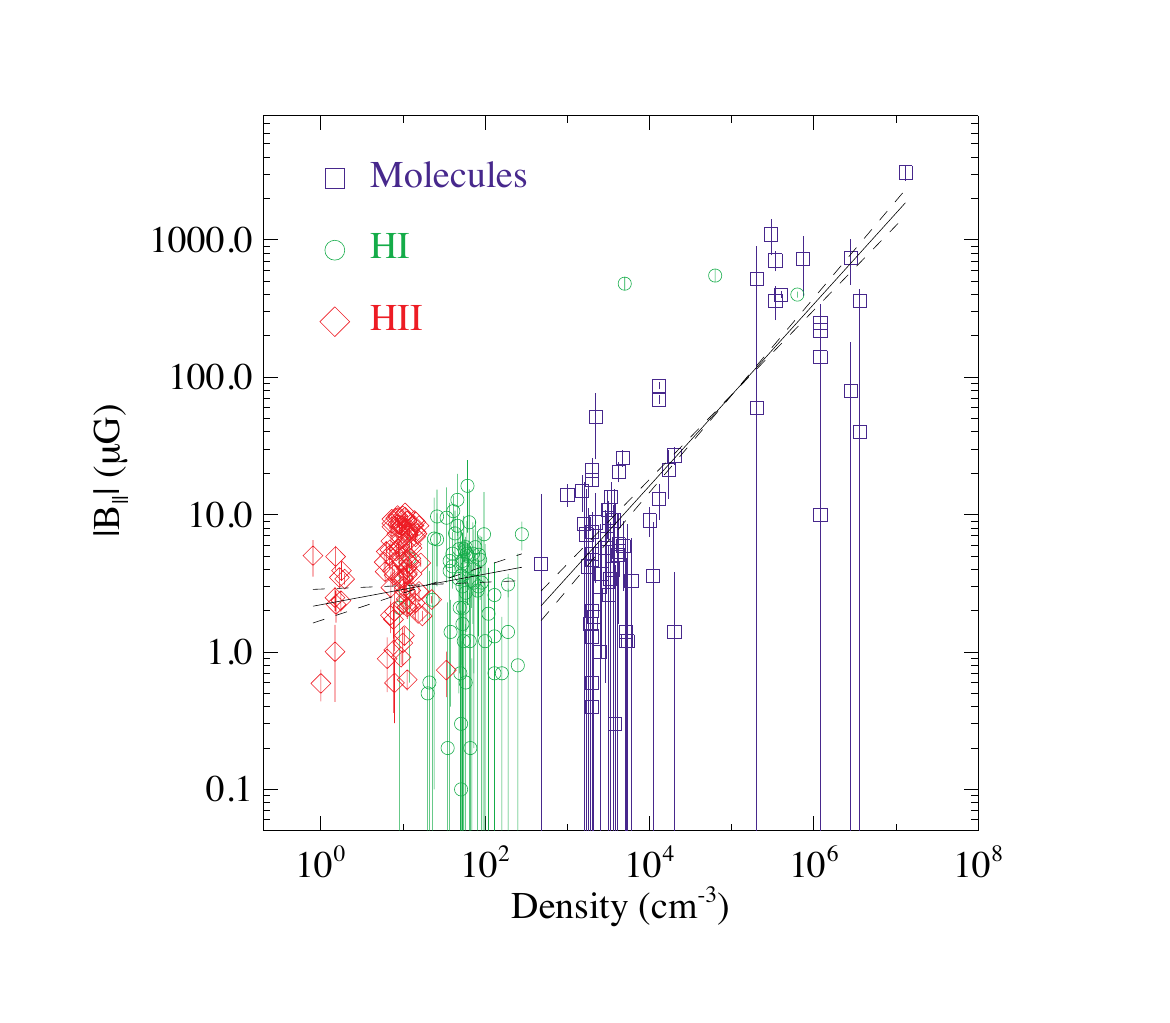}
\vspace{1cm}
\caption{\small{$B_{||}$ vs. $n$ covering nine orders of magnitude in density, using data on \HII\ regions, neutral hydrogen clouds and dense molecular cores. The {\sc Hi} and molecular data are from \protect\citet{cwh+10} and the data from \HII\ regions are our data. Separate best fits to the data (black lines) above and below $n =$ 480 cm$^{-3}$ are shown, along with dashed lines indicating the slopes of 1$\sigma$ greater and less than the least squares best fit.}}
\label{figure4}
\end{center}
\end{figure}

\begin{figure}
\begin{center}
\figurenum{5}
\includegraphics[scale=0.35]{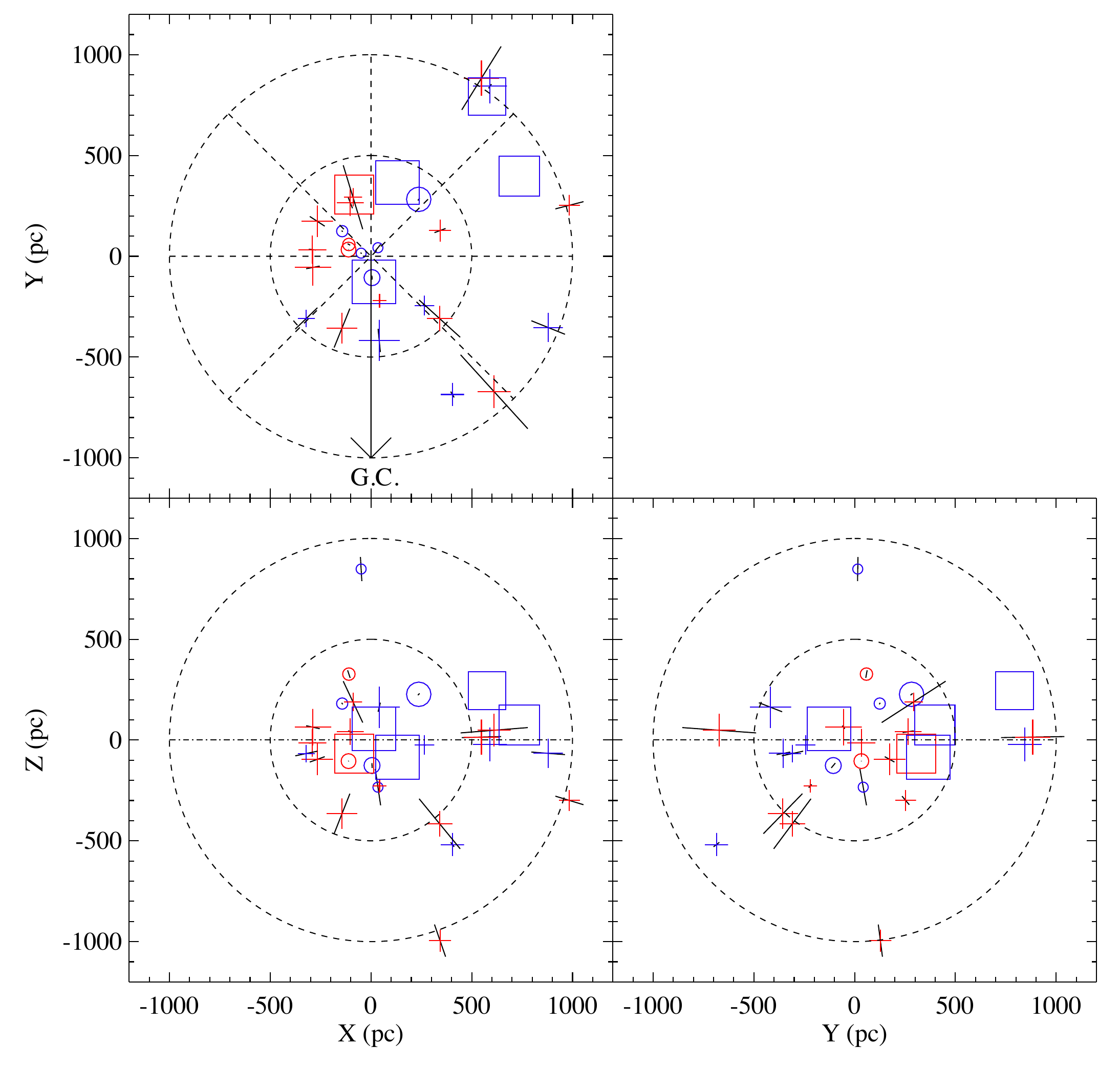}
\vspace{1cm}
\caption{\small{Orientation of the magnetic field in the solar neighbourhood, shown in three orthogonal projections. The X-Y plane is parallel and the Z-axis is perpendicular to the Galactic plane. The Sun lies at the origin. Galactic longitude increases from zero (solid line) in a counterclockwise direction (top left panel). Blue and Red symbols indicate magnetic fields oriented away from and towards the earth, respectively. The square symbols represent our data for \HII\ regions. The pulsars with parallax distance measurements \protect\citep{gtwr86,bmk+90,ktr94,bbgt02,dlrm03,btgg03,ccv+04,cbv+09,dtbr09,vbc+09} are represented by crosses.  Circles indicate sources for which the rms uncertainty in $B_{||}$ exceeds the magnitude of $B_{||}$. The size of each symbol represents the magnitude of $B_{||}$ and the black lines indicate the rms uncertainty in the distances to pulsars measured by the annual parallax method.} }
\label{figure5}
\end{center}
\end{figure}

\end{document}